\documentclass[12pt,a4paper]{article}
\usepackage{graphicx}
\usepackage{dcolumn}
\usepackage{bm}
\usepackage{amsmath,amsfonts,amssymb,setspace}
\usepackage{slashed}
\usepackage{braket,xcolor}
\usepackage{verbatim}
\usepackage{caption}
\usepackage{subcaption}
\usepackage{multirow}
\usepackage{amsfonts}
\usepackage[utf8]{inputenc}
\usepackage{setspace, hyperref}
\usepackage{cite}

\def \l{\lambda}
\def \L{\Lambda}

\def \d{\delta}
\def \o{\omega}
\def \O{\Omega}

\hypersetup{colorlinks=false, linkcolor=blue, citecolor=red}
\begin{document}
	
	\topmargin 0pt
	\oddsidemargin 0mm
	\def\be{\begin{equation}}
	\def\ee{\end{equation}}
	\def\bea{\begin{eqnarray}}
	\def\eea{\end{eqnarray}}
	\def\ba{\begin{array}}
		\def\ea{\end{array}}
	\def\ben{\begin{enumerate}}
		\def\een{\end{enumerate}}
	\def\nab{\bigtriangledown}
	\def\tpi{\tilde\Phi}
	\def\nnu{\nonumber}
	\newcommand{\eqn}[1]{(\ref{#1})}
	
	\newcommand{\vs}[1]{\vspace{#1 mm}}
	\newcommand{\dsl}{\pa \kern-0.5em /} 
	\def\a{\alpha}
	\def\b{\beta}
	\def\g{\gamma}\def\G{\Gamma}
	\def\d{\delta}\def\D{\Delta}
	\def\ep{\epsilon}
	\def\et{\eta}
	\def\z{\zeta}
	\def\t{\theta}\def\T{\Theta}
	\def\l{\lambda}\def\L{\Lambda}
	\def\m{\mu}
	\def\f{\phi}\def\F{\Phi}
	\def\n{\nu}
	\def\p{\psi}\def\P{\Psi}
	\def\r{\rho}
	\def\s{\sigma}\def\S{\Sigma}
	\def\ta{\tau}
	\def\x{\chi}
	\def\o{\omega}\def\O{\Omega}
	\def\k{\kappa}
	\def\pa {\partial}
	\def\ov{\over}
	\def\nn{\nonumber\\}
	\def\ud{\underline}
	\def\qq{$Q{\bar Q}$}
	
	\onehalfspacing
	\parskip 0.1in
	\begin{flushright}
		%
	\end{flushright}
	\begin{center}
		{\Large{\bf Islands and complexity of eternal black hole and radiation subsystems for a doubly holographic model}}
		
		\vs{10}
		
		{Aranya Bhattacharya${}^{a,b}$\footnote{aranya.bhattacharya@saha.ac.in}, Arpan Bhattacharyya${}^{c}$\footnote{abhattacharyya@iitgn.ac.in}, Pratik Nandy${}^{d}$\footnote{pratiknandy@iisc.ac.in}, Ayan K. Patra${}^{a,b}$\footnote{ayan.patra@saha.ac.in}} 
		
		\vskip 0.3in
		
		{\it ${}^{a}$Saha Institute of Nuclear Physics \\ 1/AF Bidhannagar, Kolkata 700064, India}
		\vskip .5mm
		{\it ${}^{b}$Homi Bhabha National Institute\\
			Training School Complex, Anushakti Nagar, Mumbai 400085, India}
		\vskip .5mm
		{\it ${}^{c}$Indian Institute of Technology, Gandhinagar, Gujarat-382355, India}\vskip .5mm
		{\it ${}^{d}$Centre for High Energy Physics,\\ Indian Institute of Science, C.V. Raman Avenue, Bangalore, India.}\vskip .5mm

		
		
		
		
		
	\end{center}

	\begin{abstract}
		We study the entanglement islands and subsystem volume complexity corresponding to the left/ right  entanglement of a conformal defect in $d$-dimensions in Randall-Sundrum (RS) braneworld model with subcritical tension brane. The left and right modes of the defect mimic the eternal black hole and radiation system respectively. Hence the entanglement entropy between the two follows an eternal black hole Page curve which is unitarity compatible. We compute the volumes corresponding to the left and right branes with preferred Ryu-Takanayagi (RT) surfaces at different times, which provide a probe of the subregion complexity of the black hole and the radiation states respectively. An interesting jump in volume is found at Page time, where the entanglement curve is saturated due to the inclusion of the island surfaces. We explain various possibilities of this phase transition in complexity at Page time and argue how these results match with a covariant proposal qualitatively.
	\end{abstract}
	\newpage
	\tableofcontents
	\newpage
	\section{Introduction:}
	The black hole information paradox has been one of the central and long-standing problems of theoretical physics\cite{Hawking:1974sw,Hawking:1976ra, Almheiri:2020cfm, Raju:2020smc,Harlow:2014yka}. One of the crucial concern regarding this is to understand the entanglement between the black hole and radiation. Particularly, to state that the black hole plus radiation acts as a unitary quantum system, one needs to get the time evolution of entanglement entropy to follow the Page curve (which is a characteristic feature of the unitary quantum system)\cite{Page:1993wv,Page:2013dx}. In the last couple of years, this problem has been somewhat resolved courtesy of understanding geometric entanglement and quantum extremal surface\cite{Almheiri:2019hni,Almheiri:2019psf,Almheiri:2019psy,Penington:2019kki}. The resolution came from the quantum corrected version of the holographic entanglement entropy proposal \cite{Ryu:2006bv, Ryu:2006ef, Hubeny:2007xt} known as the quantum extremal surfaces\cite{Engelhardt:2014gca}. The main role was played by new bulk regions called islands being included in the entanglement wedge of the radiation starting from the Page time, resulting in bending of the growing entanglement curve.\par
	The essence of these computations is as follows. One typically assumes that the Hawking radiation is being absorbed by a non-gravitational bath coupled with the asymptotic boundary of the gravitational system containing the black hole. Then if one considers a subregion $A$ of this bath, the entanglement entropy for this subregion is given by, 
	\begin{align} \label{eq1}
	S_{\mathrm{EE}}(\rm{A})=\mathrm{min} \bigg\{{\mathrm{ext}\atop{\mathrm{islands}}} \Big( S_{\rm{QFT}}(A \cup \rm{islands})+\frac{A(\partial (islands))}{4\, G_{N}}\Big) \bigg\}
	\end{align}
	The equation (\ref{eq1}) takes into account the entanglement entropy of quantum fields of radiation subregion $A$ together with the entanglement entropy of the gravitating subregions, termed as \textit{islands},  so that the entire functional gets minimized. At initial times, for an evaporating black hole, (\ref{eq1}) is minimized without any islands
	and result matches with the Hawking’s evaluation of the entropy. But, as time grows, island contribution dominates and it appears as a new saddle point while one minimizes (\ref{eq1}) at late times. This is due to the fact that the quanta of Hawking radiation shares an huge amount of entanglement with the quantum fields behind the horizon. At this point, the entropy is controlled by the black
	hole entropy, which appears in the second term in (\ref{eq1}), and in this way, (\ref{eq1}) produces the expected Page curve. The computation of the Page curve has been done for a variety of situations\cite{Penington:2019npb,Almheiri:2019psf,Almheiri:2019hni, Sully:2020pza,Chen:2019iro,Anegawa:2020ezn,Balasubramanian:2020hfs,Gautason:2020tmk,Hartman:2020swn,Hollowood:2020cou,Alishahiha:2020qza,Almheiri:2019psy,Rozali:2019day,Hashimoto:2020cas,Geng:2020qvw,Bak:2020enw,Li:2020ceg,Chandrasekaran:2020qtn,Almheiri:2020cfm,Hollowood:2020kvk,Bousso:2019ykv,Akers:2019nfi,Chen:2020wiq,Kim:2020cds,Verlinde:2020upt,Liu:2020gnp,Bousso:2020kmy,Balasubramanian:2020coy,Chen:2020jvn,Stanford:2020wkf,Cooper:2018cmb,Marolf:2020xie,Hartman:2020khs,Giddings:2020yes,Chen:2020tes,VanRaamsdonk:2020tlr,Sybesma:2020fxg,Balasubramanian:2020xqf,Ling:2020laa,Jian:2020qpp,Chen:2019uhq,Chen:2020hmv,Chen:2020uac,Basak:2020aaa,Kawabata:2021hac,Akal:2020ujg,Matsuo:2020ypv,Krishnan:2021faa,Krishnan:2020fer,Caceres:2020jcn, Geng:2021wcq,Manu:2020tty, Choudhury:2020hil, Anderson:2021vof}.\footnote{This list is by no means complete. Interested readers are encouraged to look into the references and citations of these papers.}\par
	In the case of an evaporating black hole, the curve comes down to zero after bending down at Page time.
	On the other hand, for the eternal black hole, the entanglement growth stops at Page time, and the curve saturates. Although the Page curve has been reproduced successfully using the idea of the islands, the full understanding of the physics is nowhere near completion. For example, we do not yet know the actual time evolution of the black hole and the radiation states. The exact physics that causes the emergence of islands is a hard nut to crack, particularly from the field theory side. 
	
	Qualitatively, for the eternal two-sided black hole, the degrees of freedom is bounded by $2 S_{BH}$ where $S_{BH}$ is the Bekenstein-Hawking entropy of the black hole. This is a constant for the eternal black hole and acts as an upper bound of the entanglement entropy. The version of Hawking's paradox for these black holes previously had an ever-growing entropy curve, and it did not stop at $2 S_{BH}$. This acts particularly as a contradiction in understanding the entanglement entropy as the fine-grained entropy. The existence of the islands solves this problem. The islands are purely geometric regions appearing in the entanglement wedge of the radiation subregion starting from the Page time. Therefore, to build a full understanding of the fine-grained entropy, it is important to look for islands from a field-theoretic scenario. Although it is largely believed that the islands capture the effects of quantum error correction\cite{Harlow:2014yka} in the holographic geometries, it is not yet understood fully from a field theory perspective. One does not know exactly how the black hole and the radiation state evolve with time. Hence, the islands appearing in the gravitational computations do not solve the information paradox of the black holes completely, and most of the physics, like understanding the reason for the appearance of islands or the exact evolution of the states of the black hole, is yet to understand. 
	
	Along these lines, another quantum information-theoretic quantity known as complexity has been shown much interest by the community in the last few years. It measures the difficulty of preparing a quantum state using a set of available quantum gates. In the bulk proposals for the complexity of a pure state, one needs to compute maximal volume slice (CV) or the classical action of the causal (Wheeler-de Witt) patch of the slice (CA)\cite{Susskind:2014moa,Susskind:2014rva,Brown:2015bva}. From a computational perspective, the complexity is lower bounded by the geodesic distance in certain manifold \cite{nielsen2006optimal, Nielsen_2006} and it was found that in view of counting the total number of gates required to prepared a unitary operator, complexity naturally scales proportionally to volume after certain optimization \cite{Bhattacharyya:2019kvj}. On the other hand, the mixed state complexity proposal is to compute the volume between the boundary subregion corresponding to the mixed state and the bulk Ryu-Takayanagi (RT) surface \cite{Alishahiha:2015rta}. This is known as the holographic subregion complexity which has been argued to be dual to the complexity of purification for the mixed state.\footnote{For details see \cite{Caceres:2019pgf,Auzzi:2019vyh,Auzzi:2019mah,Auzzi:2019fnp,Camargo:2020yfv,Ruan:2020vze,Jang:2020cbm} and references therein.} In previous work \cite{Bhattacharya:2020uun}, the complexity of radiation state for an evaporating black hole-radiation system was computed in a multiboundary wormhole model, and a subregion complexity curve corresponding to the entanglement curve was suggested. In this work, we compute respective volumes for a model where the entanglement curve is the Page curve for an eternal black hole. The basic goal is to understand the time evolution of the complexity of the radiation and the eternal black hole state in the process. 
	
	We work with a Karch-Randall (KR) brane model (a particular variant of the Randall-Sundrum (RS) braneworlds) that was studied in \cite{Geng:2020qvw, Geng:2020fxl}.\footnote{It is worth noting that in their paper, the authors argued that for a gravitating bath, one does not get a Page curve and instead end up with a constant entanglement curve. Nevertheless, in a part of their paper, the authors also show that one can get a Page curve if the entanglement between the left and right modes of the defect, where the two gravitating branes meet, is computed.} We compare the left and right modes of a conformal defect connecting two branes (one is gravitational that extends in bulk, and another one is non-gravitational that is fixed at the conformal boundary) with the eternal black hole and radiation combination. The corresponding degrees of freedom are situated on the two branes. We compute the volumes dual to the left and right branes at different times. There is indeed an appearance of island surfaces in the bulk region between the two branes. As a result, the entanglement between the left and right modes follows a Page curve typical of the entanglement between the eternal black hole and radiation. Motivated by this fact, in our model, we consider the left brane to be the analogue of the eternal black hole, whereas the right modes play the role of the radiation. We find that there are jumps and dip in volume at Page time corresponding to the right and left branes, respectively. We try to understand this jump or dip from the perspective of purification of certain Hawking modes when a new region is included in the entanglement wedge of the radiation (right modes) subsystem. It is important to keep in mind that we work with an entanglement Page curve, and the bath (/radiation) region is considered a non-gravitating one. Therefore, the whole idea is to understand the complexity of the eternal black hole and the radiation states with evolving time where the crucial physical phenomenon happens at Page time. 
	
	The rest of the paper is organized as follows. In Section \ref{Review}, we review the model in some details and discuss the Page curves of entanglement for $d=3$ and $d=4$. We review the area computations of the Hartman-Maldacena and Island surfaces as well. In Section \ref{HSC}, we compute the subregion volumes corresponding to the left and the right modes of the defect by computing volumes between the RT surface and the corresponding brane (left and right). In Section \ref{Conclusion}, we discuss our results and implications. 
	
	\section{Review of Left/Right Entanglement in RS brane model:}\label{Review}
	
	\begin{figure}
		\centering
		\includegraphics[scale=0.20]{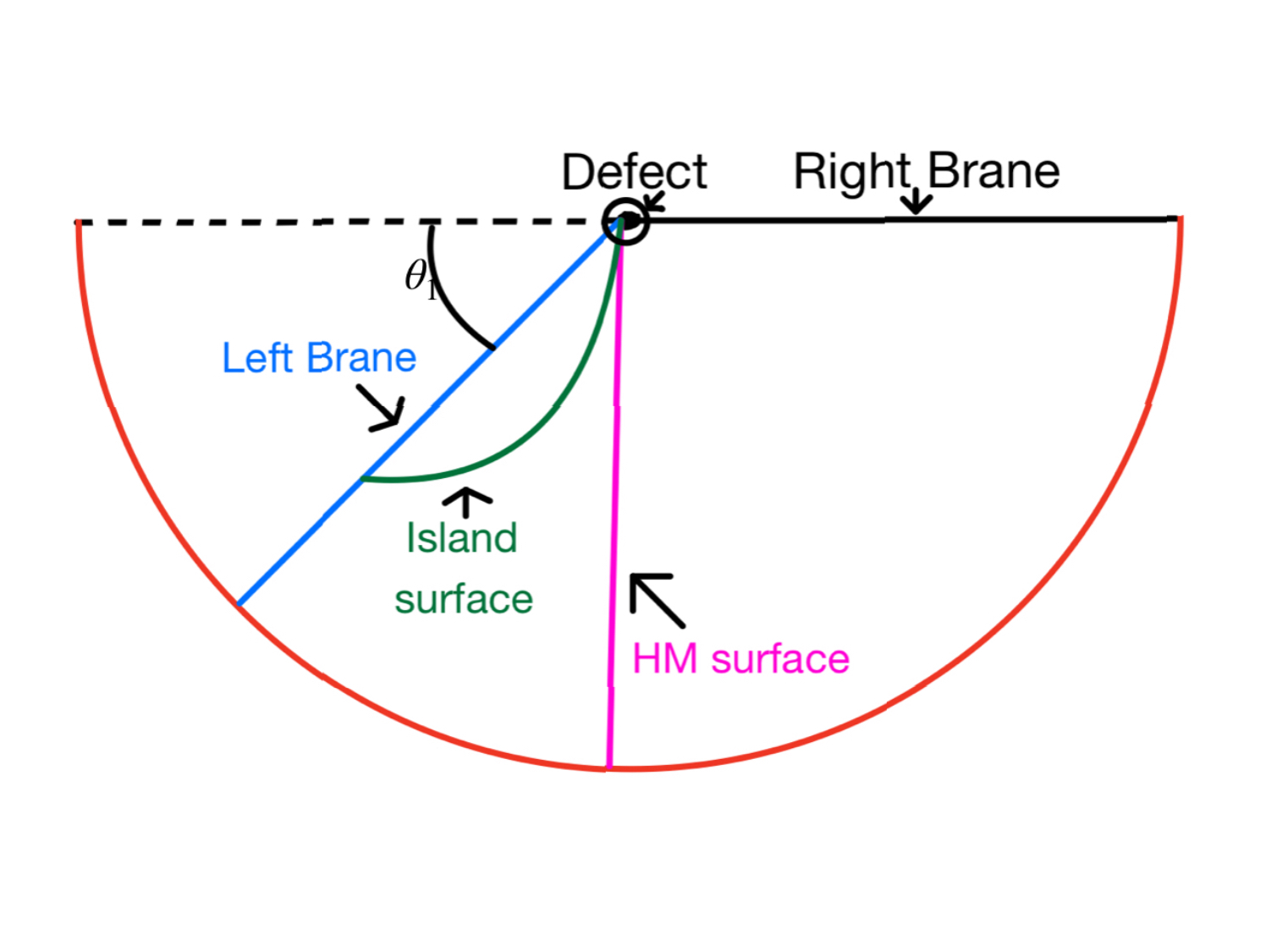}
		\caption{The braneworld model and the preferred RT surfaces before (HM surface) and after (island surface) Page time.}
		\label{basicmodel}
	\end{figure}

	In this section, we review the model and reproduce the main results we use in our work from \cite{Geng:2020fxl}. The basic goal is to model an eternal black hole and radiation system. We consider the radiation region to be non-gravitating. However, as we will argue in conclusion, effectively, this paper's main results will not change much even if a gravitating bath is considered as far as we restrict to compute the entanglement between the left and the right brane with a pre-fixed bath region. In this model, the RS braneworld framework plays a crucial role. We have two branes (Figure \ref{basicmodel}), one of which is in the bulk of a $(d+1)$-dimensional spacetime. We call it left brane or physical brane or gravitating brane interchangeably, whereas the other one is placed at the conformal boundary of the bulk, which is non-gravitating and we refer it to right brane or bath brane. The point at which these two branes meet is a conformal defect. To make the right brane gravitating, we could take it in bulk as well. We will not consider such a situation in this paper. However, we will briefly comment on this issue at the conclusion. Now, the degrees of freedom corresponding to this conformal defect is situated on both the branes. The metric in bulk is considered the metric of a $(d+1)$-dimensional black string. Initially, we distinguish the defect's left and right modes by assuming them to be situated on the left (gravitating) and right (non-gravitating) branes respectively. Then we compute the entanglement between the left and the right modes. The way to do this holographically is to find the candidate RT surfaces that start from the defect and then pick the one with the minimal area among them, measuring the entanglement entropy. But from the perspective of the defect, there is no strict constraint that can restrict the effective left modes to be situated only on the left brane and the same for the right modes.\par
	
	Therefore, we will treat all the surfaces starting from the defect and ending either on one of the branes or another defect on the other side of the black string metric as our candidate RT surfaces. As mentioned in \cite{Geng:2020qvw,Geng:2020fxl,Krishnan:2020oun}, there are two candidates for RT surfaces that are typically present in these scenarios (as shown in Figure \ref{basicmodel}). The first one is the well-known Hartman-Maldacena (HM) surface that starts from the defect and ends on the copy of the defect on the other side of the black hole. This grows linearly with time. Another candidate RT surface that came into the picture due to the concept of the island is the RT surface starting from the defect and ending on the left brane. We will call it as the island surface. When the island surface area is less than that of the HM surface, some of the modes on the left brane effectively become right modes. This can also be understood from the boundary perspective as those modes on the left brane become accessible to the right brane. The point on the left brane where the island surface ends is known as the critical anchor. In other way, we shoot surfaces from the defect and look for minimal surface ending on the left brane. It so happens that when both the HM and the island surfaces are compared, starting from a particular timescale, the island surfaces become smaller than the growing HM surface (the island surface remains constant throughout the time). This change of preferred RT surface also depends critically on the angle the left brane is placed in the bulk with respect to the conformal boundary. It is shown in \cite{Geng:2020fxl} that for only $\theta_{1}$ values upto some dimension dependent constant ($\theta_{C}=$ critical angle)\footnote{more precisely, the left/right Page curve is there for $\theta$ value upto $\theta_{P}$, which is a little less than the critical angle. In our case, the brane angles we consider are less than $\theta_{P}$ and hence $\theta_{1}<\theta_{P}<\theta_{C}$. Hence, the Page curve is there for all the angles considered in this paper.}, the competition is there. Whenever $\theta_{1}$ exceeds $\theta_{C}$, the competition is gone, and some other surfaces known as the tiny islands become the minimal RT for all times (entanglement becomes constant). It is also important to note that $\theta_{1}$ controls the strength of the gravity on the left brane. So, it seems that to get this competition or rather a Page curve, there is some bound on the strength of the gravity. Nevertheless, we will be taking three constant $\theta_{1}$ values for our purpose, all of which will be less than $\theta_{C}$. Our ultimate aim is to find the complexity curve for the situation where there is a Page curve for the entanglement between the eternal black hole and radiation. In the following subsections, we discuss the computations of the HM surface, island surface and the critical anchor that specifies the island surface for us.

	\subsection{Hartman-Maldacena and Island Areas: Revisited}
	In this section we briefly review the area of the Hartman-Maldacena (HM) and island surfaces\cite{Geng:2020fxl}. As already mentioned in \cite{Geng:2020fxl} we consider the $AdS_{d+1}$ black string metric
	\begin{align}
	ds^2 = \frac{1}{u^{2}\sin^2\mu} \bigg[ - h(u) dt^2 + \frac{d u^2}{h(u)} + d \vec{x}^2 + u^2 d \mu^2 \bigg],~~~~~ h(u)=1-\frac{u^{d-1}}{u_h^{d-1}},
	\end{align}
	to be the bulk metric. Here $u>0$ is the radial direction, $0\leq\mu<2\pi$ is the angular coordinate and $\vec{x}$ is $(d-2)$ orthogonal directions.

	\subsubsection{Hartman-Maldacena Area:}
	The Hartman-Maldacena surface is located at $\mu=\frac{\pi}{2}$\footnote{ see \cite{Geng:2020fxl} for more details.}. Thus we only need to minimize the area functional at $\mu=\frac{\pi}{2}$,
	\begin{equation}
	\mathcal{A}=\int dt \mathcal{L}
	\end{equation}
	with the Lagrangian, 
	\begin{equation}
	\mathcal{L}=u^{d-1}\sqrt{h(u)+\frac{\dot{u}^2}{h(u)}}
	\label{4}
	\end{equation}
	In $\eqref{4}$ there is no explicit time dependence in the Lagrangian thus we can write the conservation equation,
	\begin{eqnarray}
	&E=&\dot{u}\frac{\partial \mathcal{L}}{\partial{\dot{u}}}-\mathcal{L}\\
	\implies& \dot{u}=&\pm \frac{h(u)}{E}\sqrt{E^2+u^{-2(d-1)}h(u)}
	\end{eqnarray}
	where the sign is $``+"$ when $u<u_h$ and $``-"$ otherwise. The critical point $u_{\text{crit}}$ upto which we should integrate the area is determined by the relation,
	\begin{equation}
	E^2=-u_{\mathrm{crit}}^{2(1-d)}h(u_{\mathrm{crit}})
	\end{equation}  
	
	\begin{figure}
		\centering
		\begin{subfigure}[b]{0.45\textwidth}
			\centering
			\includegraphics[scale=.55]{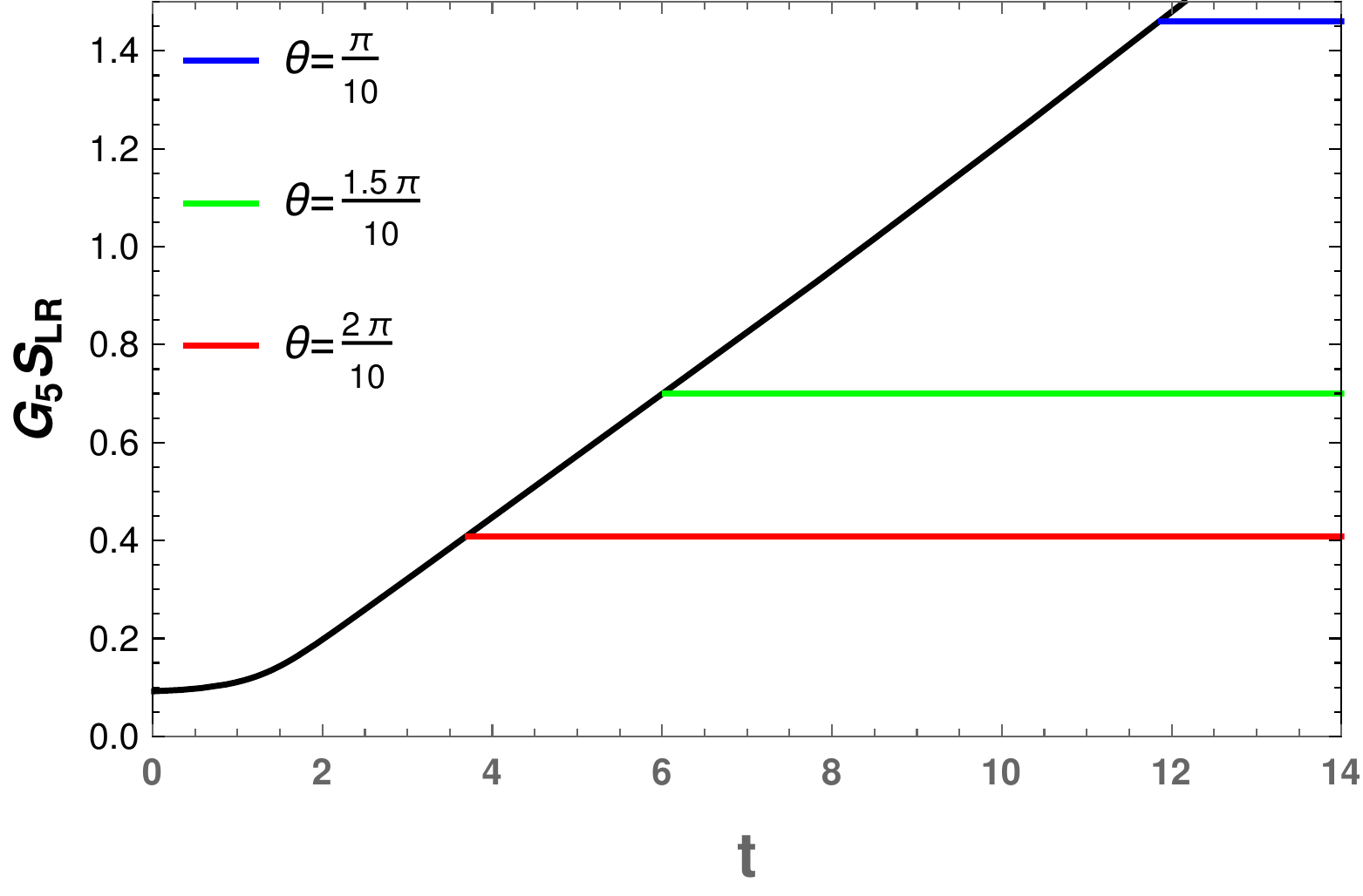}
			\caption{Page curve for $d=4$.}
			\label{areatd4}
		\end{subfigure}
		~~~~~~
		\begin{subfigure}[b]{0.45\textwidth}
			\centering
			\includegraphics[scale=.55]{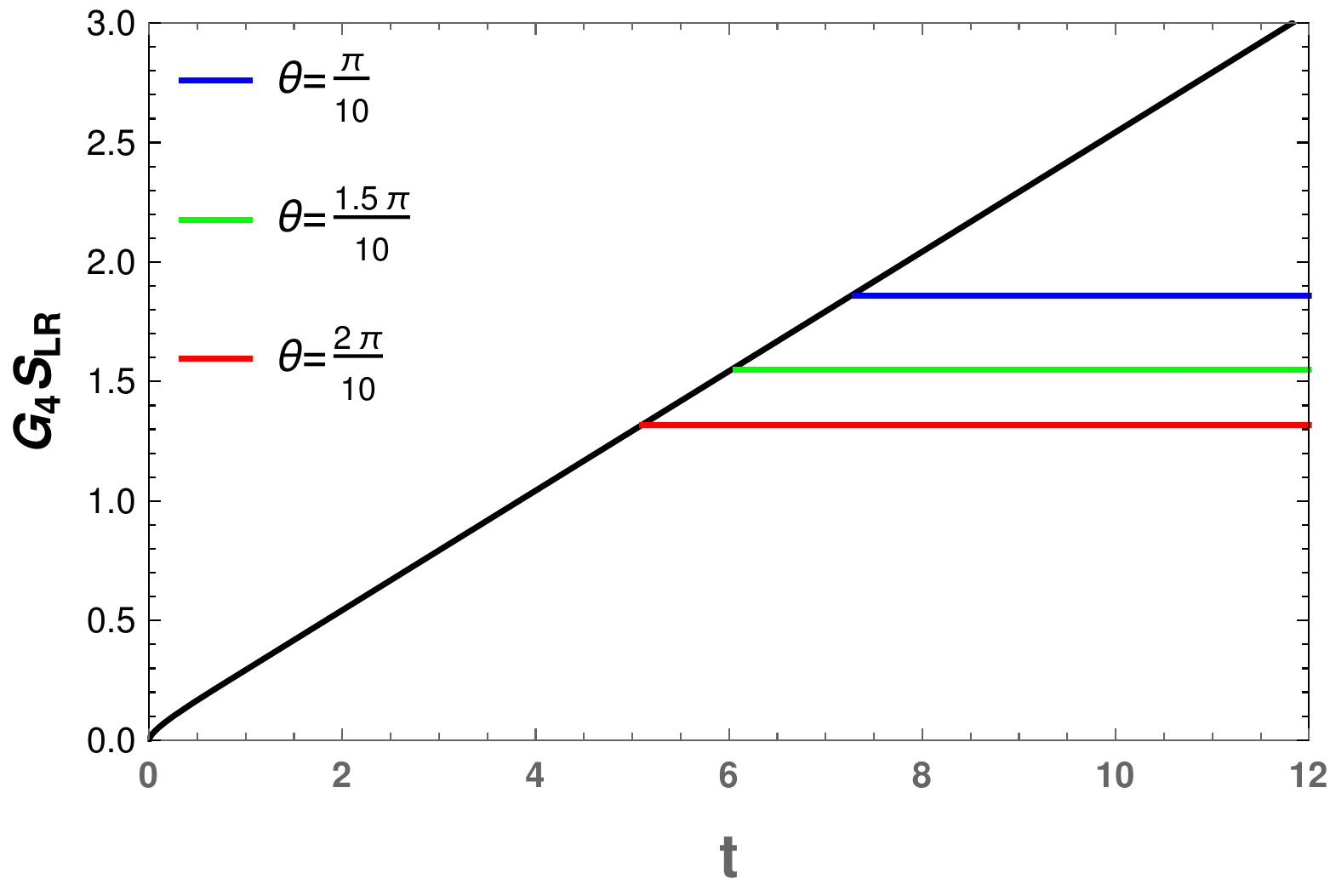} 
			\caption{Page curve for $d=3$.}
			\label{areatd3}
		\end{subfigure}
		\caption{Page curves for $d=4$ and $d=3$.}
		\label{areat}
	\end{figure}
	
	As $u\rightarrow{u_{\mathrm{crit}}}$, the boundary time as well as the HM area diverges. Therefore the minimal area is,
	\begin{equation}
	\mathcal{A}_{HM}=\lim_{\delta\rightarrow 0}\int_{\delta}^{u_\text{crit}}\frac{du}{|\dot{u}|u^{d-1}}\sqrt{-h(u)+\frac{\dot{u}^2}{h(u)}}
	\label{8}
	\end{equation}
	It can be easily verified that the area in $\eqref{8}$ contain a divergent piece when $\delta \rightarrow 0$. This divergent term is $\frac{1}{(d-2)\delta^{d-2}}$ as $\delta\rightarrow0$. Therefore we can regularise the HM area by subtracting this divergent term,
	\begin{equation}
	A^{\mathrm{reg}}_{HM}(t_{\text{diff}})=\lim_{\delta\rightarrow{0}}\left[-\frac{1}{(d-2)\delta^{d-2}}+\int_{\delta}^{u_{\mathrm{crit}}}\frac{du}{|\dot{u}|u^{d-1}}\sqrt{-h(u)+\frac{\dot{u}^2}{h(u)}}\right]
	\end{equation}
	On the other hand $t_{\text{diff}}$ is given by,
	\begin{equation}
	t_{\text{diff}}=\int_{0}^{u_{\mathrm{crit}}}{t'(u)}du={\lim_{\epsilon\rightarrow 0}}\int_{0}^{u_h-\epsilon}\frac{du}{\dot{u}}+\int_{u_h+\epsilon}^{u_{\mathrm{crit}}}\frac{du}{\dot{u}}
	\end{equation}
	By varying the $u_{\mathrm{crit}}$ we can find the regularised Hartman-Maldacena area as a function of $t_{\text{diff}}$.
	
	
	\subsubsection{Island Area:} 
	If and when there is some surface present starting from the defect and ending on one of the branes with a lesser surface area than the HM surface, it can become the preferred RT surface. We call such a surface the island surface. As shown in \cite{Geng:2020qvw, Geng:2020fxl}, the island surface ends on the brane, which is more gravitating among the two. Therefore, in our case, we search for the island surface on the left brane. The island surfaces are time-independent surface because, unlike the HM surface, the embedding function is found by solved with a timeslice and considering $u=u(\mu)$ embedding. In $d=4$ to compute the island area we need to minimize the area functional given by
	\begin{align}
	\mathcal{A} = \int \frac{d \mu}{(u \sin \mu)^3} \sqrt{u(\mu)^2 + \frac{u'(\mu)^2}{h(u)}} \label{areaf}
	\end{align}
	
	Practically what one does to find the critical anchor (the point on the left brane where the island surface ends) is to start from different points on the left brane and solving the embedding function by taking different boundary conditions $u(\theta_{1})=u_{1}/, \text{and}\, \, u^{\prime}(\theta_{1})=0$. Then the job is to check if the surface satisfying the equation of motion can reach the conformal defect at $\mu=\frac{\pi}{2}$. This is mentioned briefly in the following section \ref{criticalanc}.\par
	
	After finding out the critical anchors and hence, the island surfaces for different physical brane angles, we regularize both the HM and island surfaces. We discuss the regularization of island surface area in Appendix \ref{appa}. The HM and island areas are plotted for $d=3,4$ in Figure \ref{areat}.\footnote{In Figure \ref{areat}, we plotted $\frac{\text{Area}\left(\gamma\right)}{4}$ in the $y$ axis using the RT formula $S_{LR}=\frac{\text{Area}\left(\gamma\right)}{4G_{d+1}}$ \cite{Ryu:2006bv}. $\gamma$ is the minimal surface.} As can be seen from the figures, there is a crossover between the HM and island areas at some point in time. This time is different for different brane angles and spacetime dimension. This time is known as the Page time ($t_{\mathrm{Page}}=t_{\mathrm{Page}}(\theta_{1},d)$) where the RT surface area changes from the time dependent to the time independent one. For $d=4$, the Page times are
	$\approx 11.87$, $\approx 6.03$ and $\approx 3.71$ for $\theta{1}=\frac{\pi}{10}$, $\frac{1.5\pi}{10}$ and $\frac{2\pi}{10}$ respectively (Figure \ref{areatd4}). For $d=3$, the Page times are
	$\approx 7.3$, $\approx 6.07$ and $\approx 5.1$ for $\theta_{1}=\frac{\pi}{10}$, $\frac{1.5\pi}{10}$ and $\frac{2\pi}{10}$ respectively. For a constant $d$, $t_{\mathrm{Page}}$ decreases with increasing physical brane angle $\theta_{1}$.

	\subsubsection{Critical Anchor at $d=4$ and $d=3$:}\label{criticalanc}
	The embedding action for the black string metric in $d=4$ is given by Eq. \eqref{areaf}
	\begin{align}
	\mathcal{A} = \int_{\theta_1}^{\pi - \theta_2} \frac{d \mu}{(u \sin \mu)^3} \sqrt{u(\mu)^2 + \frac{u'(\mu)^2}{h(u)}}
	\end{align}
	
	As mentioned before, we fix the physical (left) brane at three different angles $\theta = \pi/10,~ \theta = 1.5\,\pi/10$ and $\theta = 2 \pi/10$. Extremizing the action with the boundary condition imposed as $u(m\pi/10) = u_1$ and $u'(m\pi/10) = 0$ with $m=1, 1.5, 2$ respectively, where the derivative vanishes according to the Neumann boundary condition \cite{Geng:2020fxl}. 
	
	\begin{figure}
		\centering
		\begin{subfigure}[b]{0.32\textwidth}
			\centering
			\includegraphics[width=\textwidth]{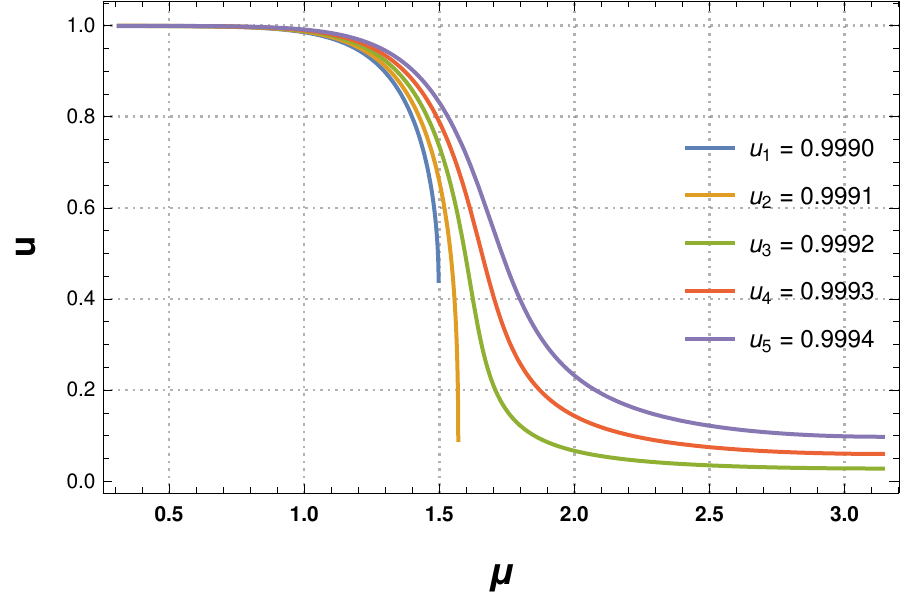}
			\caption{$\theta = \pi/10$.}
			\label{fig:pi4}
		\end{subfigure}
		\hfill
		\begin{subfigure}[b]{0.32\textwidth}
			\centering
			\includegraphics[width=\textwidth]{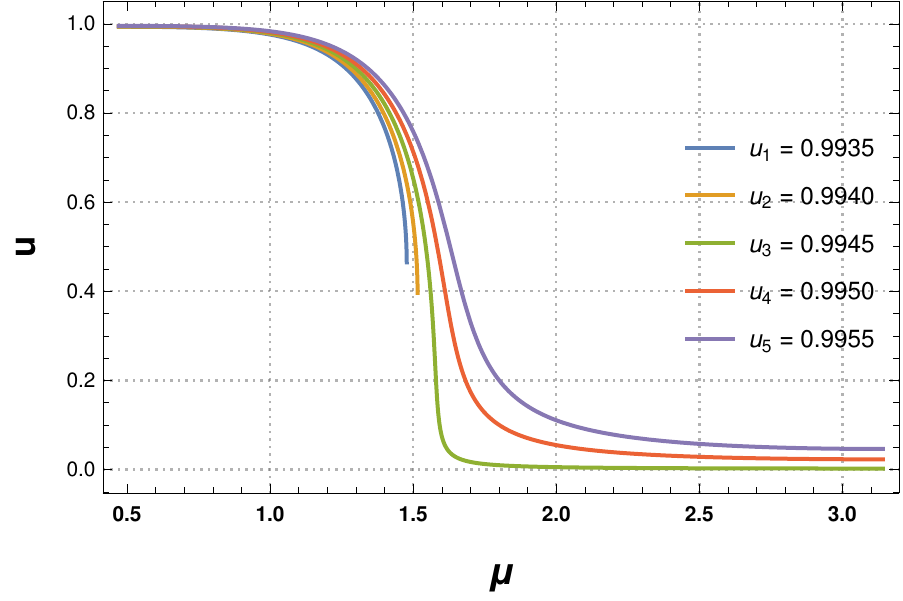}
			\caption{$\theta = 1.5\, \pi/10$.}
			\label{fig:halfpi4}
		\end{subfigure}
		\hfill
		\begin{subfigure}[b]{0.32\textwidth}
			\centering
			\includegraphics[width=\textwidth]{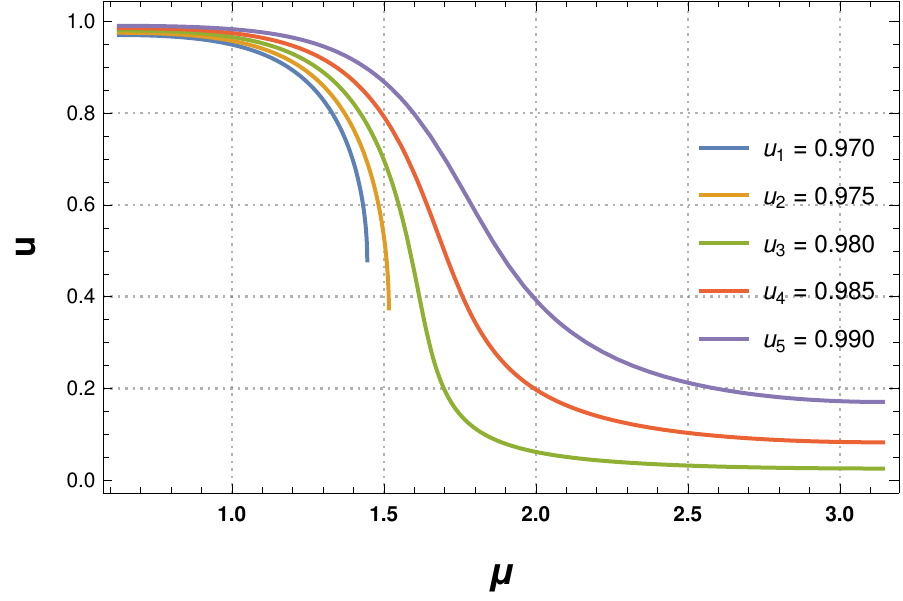}
			\caption{$\theta = 2 \pi/10$.}
			\label{fig:twopi4}
		\end{subfigure}
		\caption{Critical anchor at $d=4$ for different values of physical (left) brane angle.}
		\label{fig:d4criticalanchor}
	\end{figure}
	\begin{figure}
		\centering
		\begin{subfigure}[b]{0.32\textwidth}
			\centering
			\includegraphics[width=\textwidth]{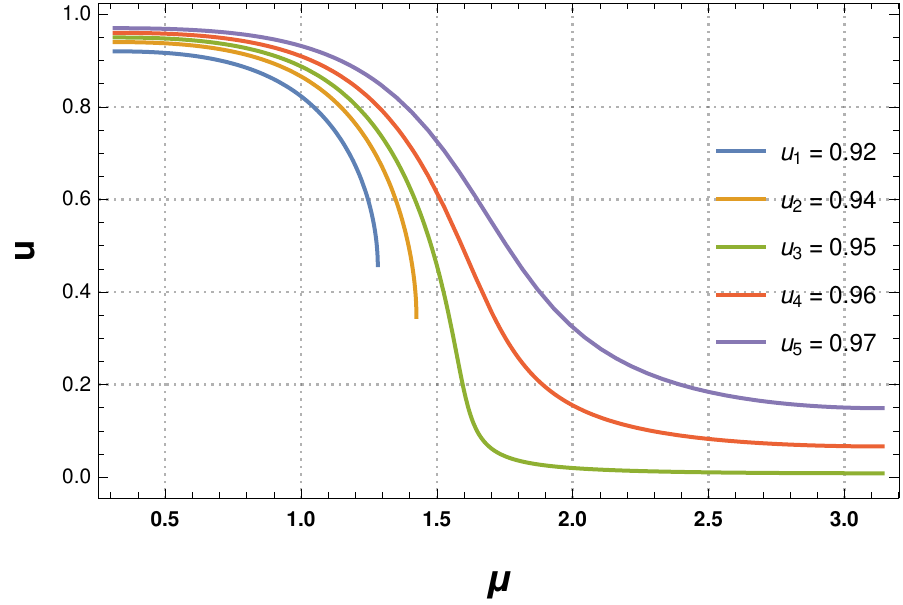}
			\caption{$\theta = \pi/10$.}
			\label{fig:pi}
		\end{subfigure}
		\hfill
		\begin{subfigure}[b]{0.32\textwidth}
			\centering
			\includegraphics[width=\textwidth]{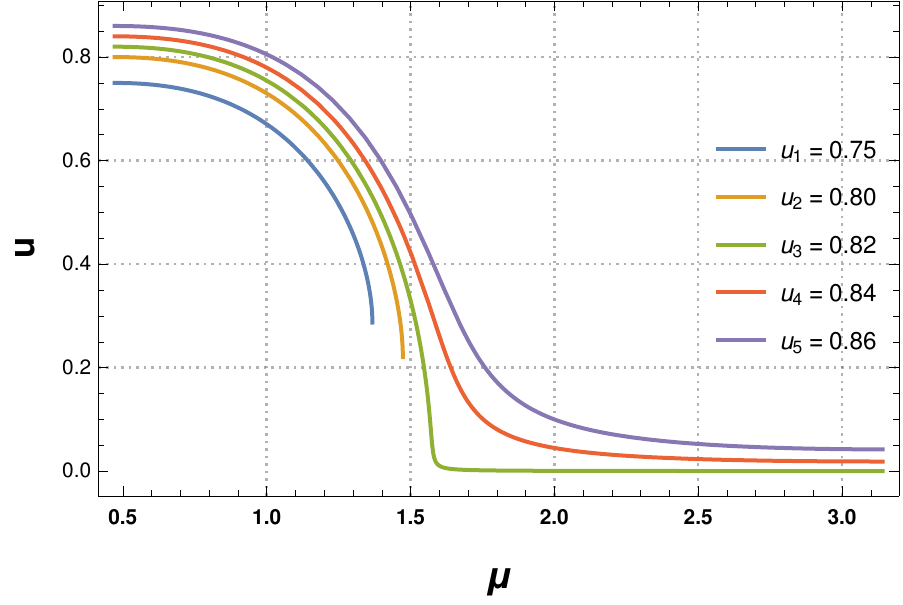}
			\caption{$\theta = 1.5\, \pi/10$.}
			\label{fig:halfpi}
		\end{subfigure}
		\hfill
		\begin{subfigure}[b]{0.32\textwidth}
			\centering
			\includegraphics[width=\textwidth]{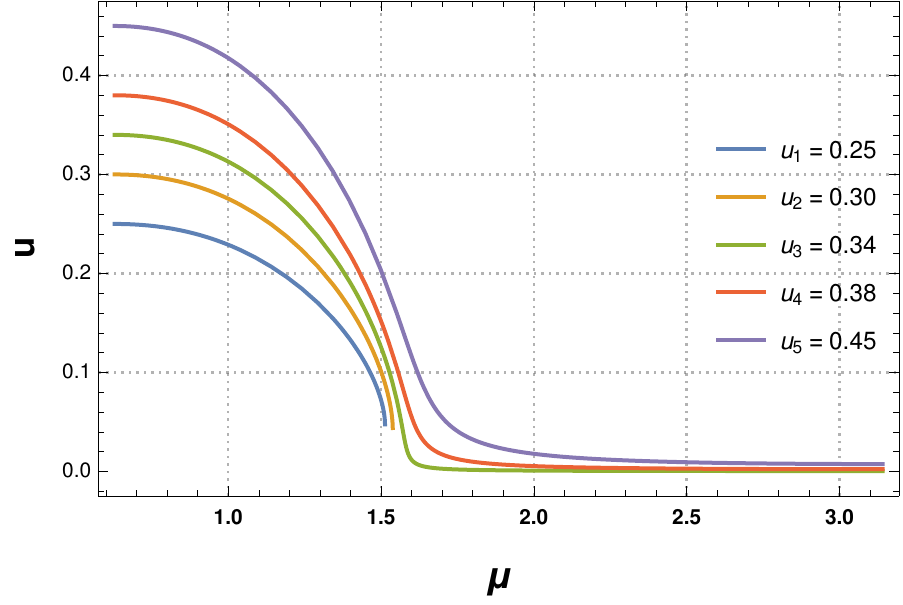}
			\caption{$\theta = 2 \pi/10$.}
			\label{fig:twopi}
		\end{subfigure}
		\caption{Critical anchor at $d=3$ for different values of physical (left) brane angle.}
		\label{fig:d3criticalanchor}
	\end{figure}
	
	We see that there exists a cut off of $u$ on the physical brane where the geodesics reach the defect (Figure \ref{fig:d4criticalanchor}). The values of the $u_{1}$ is chosen by choosing different $u_{1}$'s within a range of values and look for the surface that reaches $\mu=\frac{\pi}{2}$ (defect), but does not reach $\mu=\pi$ (conformal boundary). This happens for the orange plots in Figure \ref{fig:d4criticalanchor} (for $d=4$) and Figure \ref{fig:d3criticalanchor} (for $d=3$). The $u_{1}$ values for $d=4$ are $\approx 0.9991$, $\approx 0.9940$ and $\approx 0.975$ for $\theta_{1}=\frac{\pi}{10}$, $\frac{1.5\pi}{10}$ and $\frac{2\pi}{10}$ respectively.

	We find that while increasing the angle of the physical brane, the critical anchor value decreases. This implies as gravity gets stronger on the physical brane, the geodesics corresponding to the RT surface start from smaller numerical values, i.e., the volume of RT surfaces will decrease with the increase of gravity on the physical brane consistent with the expectation. The critical anchor's behaviour in $d=3$ will also follow the same pattern (Figure \ref{fig:d3criticalanchor}). Note that the physical brane must be placed at an angle lower than the critical angle corresponding to that specific dimension for all the cases.

	\section{Volume Subregion Complexity:}\label{HSC}
	
	In this section, we apply the proposal of\cite{Alishahiha:2015rta} to compute the volume below the RT surface at different times. The proposal introduced in \cite{Alishahiha:2015rta} states that the volume below the RT surface gives the subregion complexity of a mixed state in AdS/CFT at that given time. To be precise, the proposal deal with static scenarios. Therefore, no change in the RT surface was considered there. Given a RT surface, the proposal, therefore, states that the volume subregion complexity ($C_{A}$) for a subregion mixed state $A$ is defined as
	
	\begin{equation}\label{subcomp}
	C_{A}=\frac{V\left(\gamma_{RT(AA^{C})}\right)}{8 \pi \ell G_N},
	\end{equation}
	
	where the denominator is a particular normalization involving the bulk Newton's constant $G_N$ and the AdS length scale $\ell$, here $\gamma_{RT(AA^{C})}$ is the RT surface that divides the bulk corresponding to boundary subregions $A$ and $A^{C}$ (compliment of $A$) into two bulk subregions. In a further study, we will suppress this normalization in the denominator of eq. \eqref{subcomp} as well as the constant factor in the numerator coming from the extra dimensions. Hence, in a way, what we compute is the renormalized volume subregion complexity density. \par
	
	The covariant proposal of volume subregion complexity \cite{Agon:2018zso} for time-dependent cases is a combination of the ``Complexity$=$ Volume" proposal \cite{Susskind:2014moa, Susskind:2014rva} and Subregion complexity proposal for static cases \cite{Alishahiha:2015rta}. The proposal goes as follows. Given a time-dependent metric, we have a time-dependent RT surface, also known as the Hubeny-Rangamani-Takayanagi (HRT) surface \cite{Hubeny:2007xt}. At a given boundary time $t_0$ and a boundary subsystem $A(t_{0})$, let us say the HRT surface is $\gamma(t_{0},t)$ (here we emphasize the fact that the RT surface is generally time-dependent and it does not necessarily stay at the bulk slice $t=t_{0}$). Then according to the covariant proposal of subregion volume complexity, one should look for co-dimension one bulk slices $\Sigma_{A}(t_{0},t)$ which has boundary $\partial \Sigma_{A}(t_{0},t)= A(t_{0}) \cup \gamma(t_{0},t)$. Then there is supposed to be an infinite number of such slices $\Sigma_{A}$, and we take the one with the maximal volume. This can also be simply stated as finding the maximal volume Cauchy slice of the entanglement wedge. It is important to remember at this point that the definition of the entanglement wedge is the bulk domain of dependence, bounded by $\partial \Sigma_{A}(t_{0},t)$. This proposal involves complicated computations in general but is self-consistent and works for any subsystem $A$. This also boils down to Alishahiha's proposal \cite{Alishahiha:2015rta} if we take a time-independent scenario, where the $t$ dependence of $\Sigma_{A}(t_{0},t)$ is not there anymore, and there is a single choice of volume to compute. 
	
	\begin{equation}\label{maxvol}
	C_{A_{\mathrm{cov}}}(t)= \text{Max}_{\Sigma_{A}(t_{0},t)} \bigg[\frac{V\left(\Sigma_{A}\left(t_{0},t\right)\right)}{8\pi \ell G_{N}}\bigg].
	\end{equation}

	In our model, the situation is non-static. Hence, ideally, we should be following the covariant proposal of subregion complexity. However, as mentioned previously, technically, it is extremely nontrivial and hard to find the maximal volume slice explicitly among all possible volume slices. We, therefore, resolve to a different path. We follow Alishahiha's proposal\cite{Alishahiha:2015rta} instead of the covariant one. We consider volume below the HRT surface, which has the minimal area at a given time, to be the subregion complexity at that particular time. We therefore compute the volume corresponding to the boundary time $t_{0}$ for different times $t_{0}$. However as we argue later in the paper, the qualitative nature of our study will remain unchanged for the covariant proposal as well. With this, let us now discuss the volume computation in the model we are interested.\footnote{The covariant prescription of volume subregion complexity was used in \cite{Hernandez:2020nem}. In our proposal, we fix the time coordinate as the boundary time and compute volumes for different times in a similar spirit to what was done in \cite{Chen:2018mcc}. In a time-dependent case, it is somewhat similar to taking different snapshots taken from the boundary time and computing volumes enclosed by the HRT and the boundary subregion as seen by an observer sitting on the boundary. However, it will be interesting to study subregion complexity using maximal volumes. We hope to explore this direction in future.}
	
	As mentioned in the previous section, the left and right modes of the defect play the role of the eternal black hole and radiation respectively in our model. Therefore, the degrees of freedom contributing to the entanglement between left and right modes on the left brane decrease in number after the Page time. Similarly, the decreased degrees of freedom result in the increase in the effective degrees of freedom corresponding to the right brane. Starting from the Page time, part of the degrees of freedom on the left (gravitating) brane belong to the right modes of the defect, which results in saturation of the Page curve. In the following, we compute volumes corresponding to both the left and the right branes to get an idea about the subregion complexity of the evolving left and right modes (eternal black hole and radiation). 
	
	\begin{figure}[h!]
		\centering
		\begin{subfigure}[b]{0.45\textwidth}
			\centering
			\includegraphics[width=\textwidth]{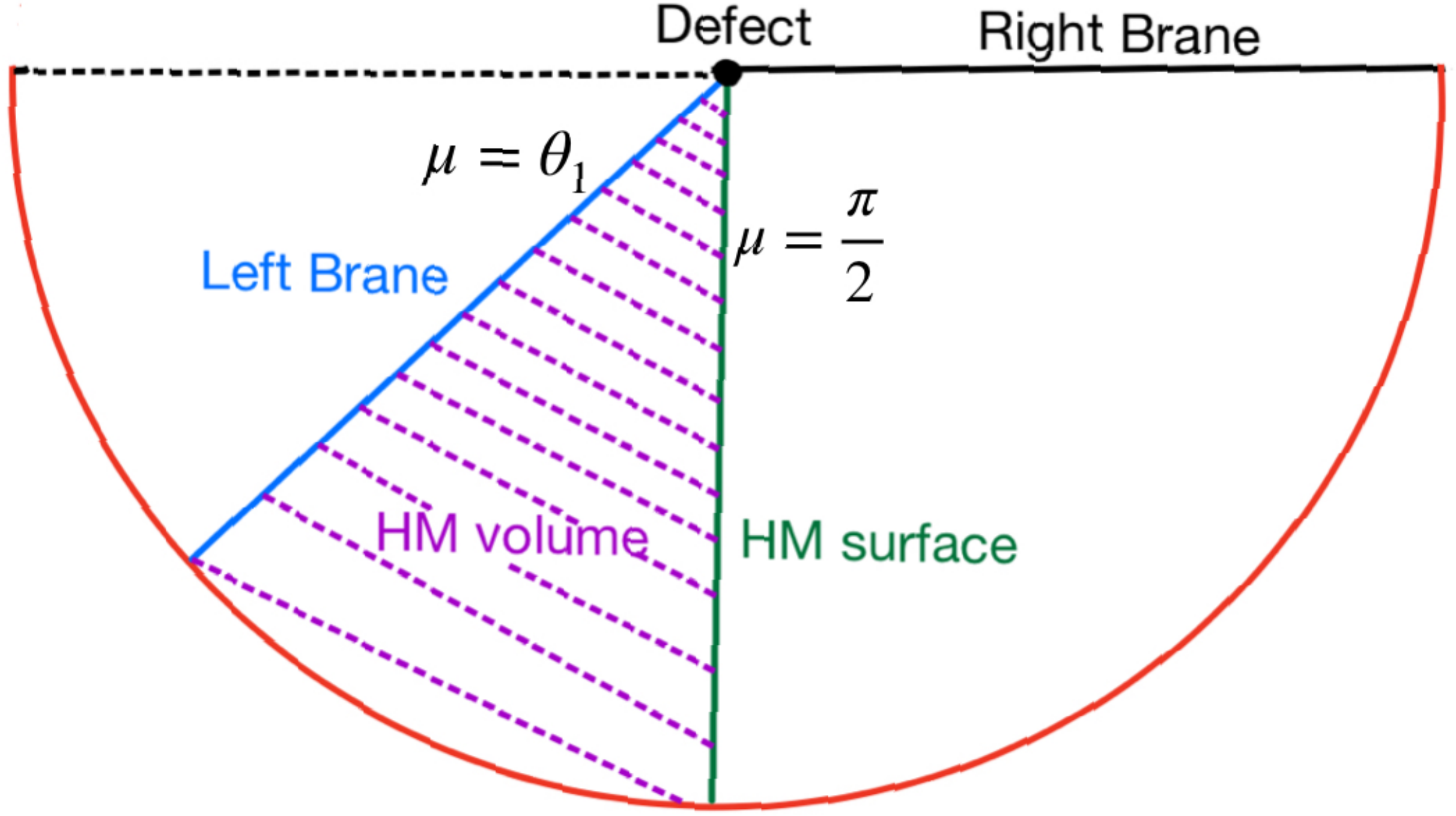}
			\caption{Volume between physical brane and HM surface.}
			\label{figHMvol}
		\end{subfigure}
		~~~~~~
		\begin{subfigure}[b]{0.45\textwidth}
			\centering
			\includegraphics[width=\textwidth]{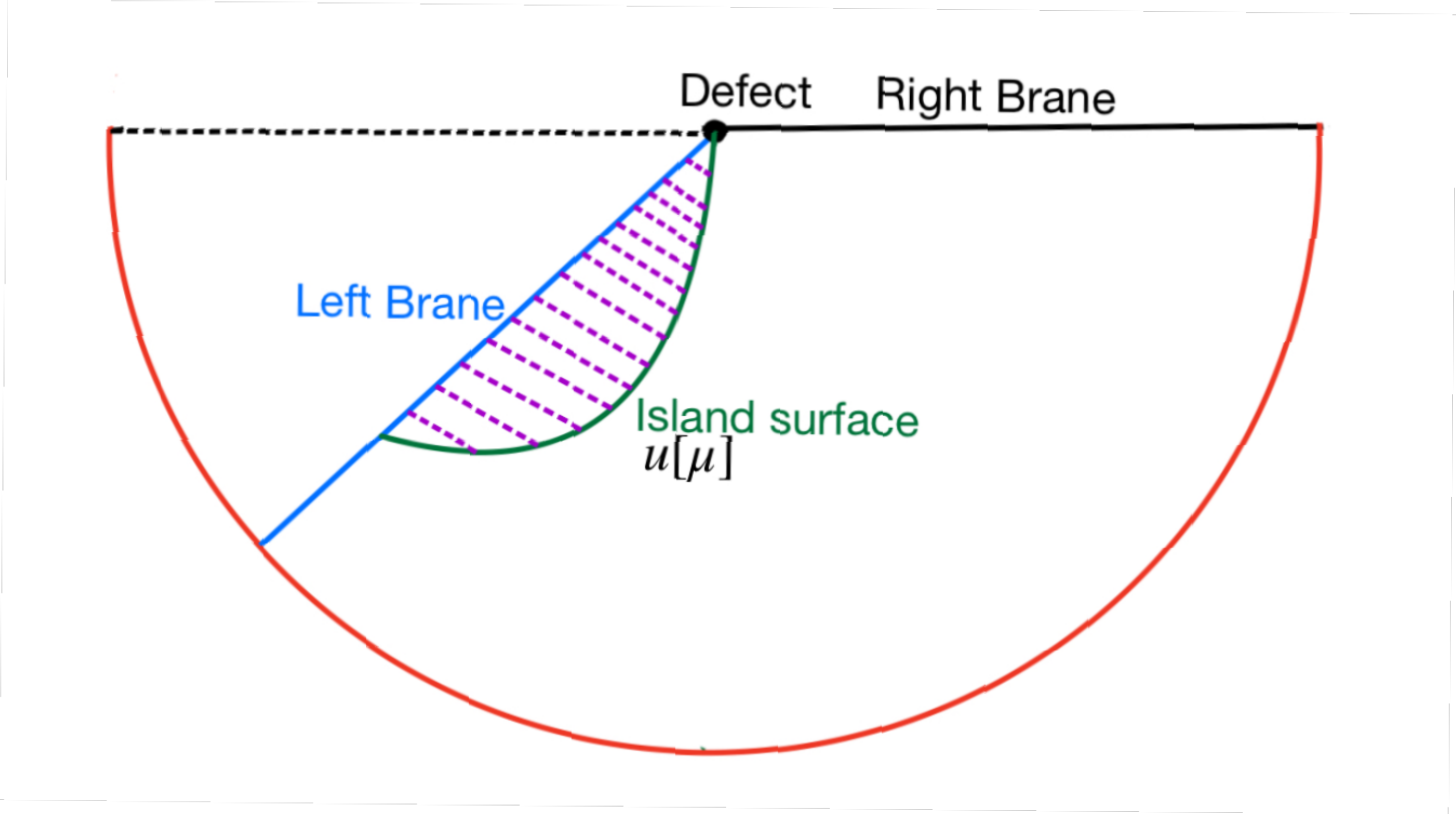}
			\caption{Volume under island on physical brane.}
			\label{figislandvol}
		\end{subfigure}
		\caption{Left (physical/gravitating) brane volumes before and after the Page time are shown. Before Page time, HM surface $u=u(t)$ is the RT surface whereas after Page time, the preferred RT surface is the Island surface $u(\mu)$ specified by the critical anchor. Correspondingly volumes on the left brane become (\ref{figHMvol}) before Page time and (\ref{figislandvol}) after page time.}
		\label{fig:leftbranevols}
	\end{figure} 
	
	\subsection{Volumes Corresponding to Left Brane (Eternal Black Hole):}\label{volleft}
	For the volume between the RT surface and the left brane, the protocol is following. We compute the volume between the HM surface and the complete left brane until Page time. After Page time, as the island surface appears to be the preferred RT surface, the subregion complexity of the left modes is given by the volume between the island surface and the left brane.
	
	\subsubsection{Hartman-Maldacena Volume:}\label{lefthmvoll}
	Here we compute the $d$-dimensional volume between HM surface and the left brane, which is the marked region in Figure (\ref{figHMvol}).
	To do that we again start with AdS$_{d+1}$ black string metric,
	\begin{equation}\label{HMmetric}
	ds^2=\frac{1}{u^2\sin^2{\mu}}\left[-h(u)dt^2+\frac{du^2}{h(u)}+d\vec{x}^2+u^2d\mu^2\right]
	\end{equation}
	where, blackening factor $h(u)=1-\frac{u^{d-1}}{u_h^{d-1}}$. As discussed in earlier section the embedding function $u(\mu,t)$ is only function of $t$ and not of $\mu$ for Hartman-Maldacena surface. Therefore, we only consider the metric for constant $\mu$ value,
	\begin{align}\label{timeslicee}
	ds^2|_{\mu=\mathrm{constant}}=&\frac{1}{u^2\sin^2\mu}\left[-h(u)dt^2+\frac{du^2}{h(u)}+d\vec{x}^2\right]\\
	=&\frac{1}{u^2\sin^2\mu}\left[\left(-h(u)+\frac{{\dot{u}}^2}{h(u)}\right)dt^2+d\vec{x}^2\right]
	\end{align}
	
	Now we compute the area for each constant $\mu$ slices and minimize it. After the minimization, we can choose the same embedding function $u(t)$ for every constant $\mu$ value. Therefore, $d$-dimensional volume should be just the area multiplied by a factor that depends on the angle of the physical (left) brane,
	

	\begin{align}
	V_{L-HM}(\theta_1, t_{\text{diff}})&= C_1\mathcal{I}(\theta_1) A^{\mathrm{reg}}_{HM}(t_{\text{diff}})\\
	\mathcal{I}({\theta_1})&=\int_{\theta_1}^{\frac{\pi}{2}}\frac{d\mu}{\sin^d\mu}\\
	A^{\mathrm{reg}}_{HM}(t_{\text{diff}})&=\lim_{\delta\rightarrow{0}}\left[-\frac{1}{(d-2)\delta^{d-2}}+\int_{\delta}^{u_{\mathrm{crit}}}\frac{du}{|\dot{u}|u^{d-1}}\sqrt{-h(u)+\frac{\dot{u}^2}{h(u)}}\right]
	\end{align}
	where $C_{1}(=\int d\vec{x})$ is the constant volume factor contributed by the transeverse directions. It is worth noting that this factor was also present but suppressed in case of the HM area.
	
	Now, we know from our entanglement study that the regularized area $A^{\mathrm{reg}}_{HM}$ is a monotonically increasing function of $t_{\text{diff}}$. Hence, asymptotically the volume $V_{L-HM}$ also goes linearly with $t_{\text{diff}}$. However, the volume depends on the angle the physical brane makes with the conformal boundary due to the $\mathcal{I}(\theta_{1})$ factor.


	
	\subsubsection{Volume after Page time:}
	Now let us focus on the island volume computation. For island the embedding function $u$ only depends on $\mu$. At $t=\text{0}$ metric is,
	\begin{equation}
	ds_{\tilde{g}}^2=\frac{1}{u^2\sin^2{\mu}}\left[\frac{du^2}{h(u)}+d\vec{x}^2+u^2d\mu^2\right]
	\end{equation}
	Therefore the volume between the island and the left brane is,
	\begin{align}
	V_{L-IS}(\theta_1)=&\int\sqrt{\tilde{g}}\,d^{d}x\\
	=& C_{1}\left(\int\frac{du\,d\mu}{\sqrt{h(u)}u^{d-1}\sin^{d}\mu}\right)\\
	=& C_{1}\lim_{\delta\rightarrow{0}}\int_{\theta_1}^{\frac{\pi}{2}}\left(\int_{\delta}^{u(\mu)}\frac{du}{\sqrt{h(u)}u^{d-1}}\right)\frac{d\mu}{\sin^d\mu},
	\label{23}
	\end{align}
	As mentioned previously, we suppress this factor to plot the subregion complexity density.
	One can check that the divergence of the island volume comes from the $u$ integration. Therefore, we only need to introduce a cut-off there. The regularized volume is,
	\begin{equation}
	V_{L-IS}^{\mathrm{reg}}(\theta_1)=\lim_{\delta\rightarrow{0}}\int_{\theta_1}^{\frac{\pi}{2}}\left(\int_{\delta}^{u(\mu)}\left(\frac{du}{\sqrt{h(u)}u^{d-1}}-\frac{1}{(d-2)\delta^{d-2}}\right)\right)\frac{d\mu}{\sin^d\mu}
	\end{equation}
	
	\begin{figure}
		\centering
		\includegraphics[scale=0.55]{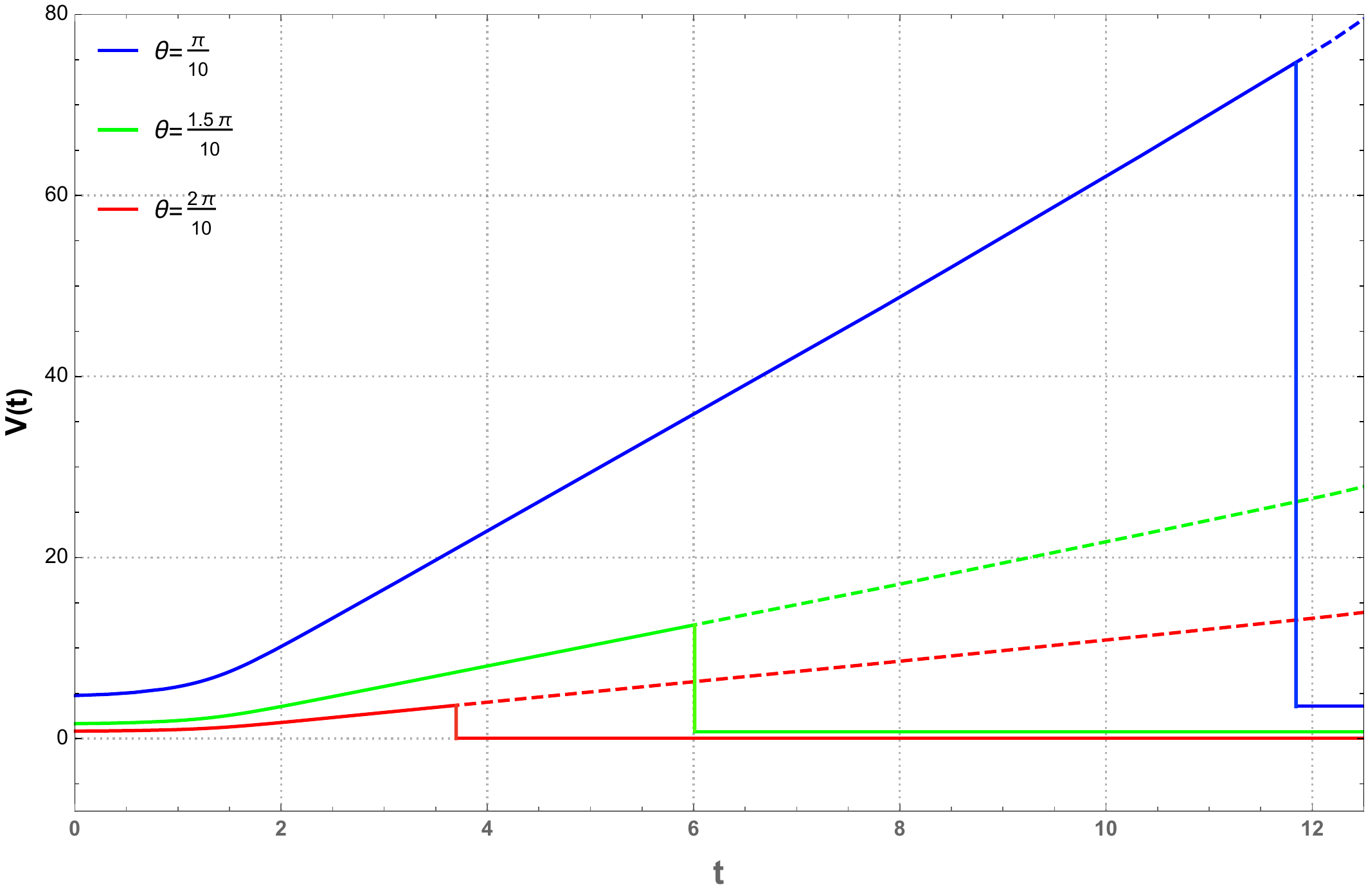}
		\caption{Volume vs time plot corresponding to the left brane for $d=4$. The blue, green and red plots correspond to the volumes for physical branes with brane angles $\theta_{1}=\frac{\pi}{10},\, \, \frac{1.5 \pi}{10} \,\, \text{and}\,\, \frac{2\pi}{10}$ respectively. For all the three cases, initially the linearly growing plots are favoured (volume dual to HM surface on left brane). After the Page time, which depends on the brane angle for each case, the constant straight line curves are favoured as island surfaces become the preferred RT surfaces.}
		\label{voltleftd4}
	\end{figure}
	
	So the divergence is of the form $-\frac{\mathcal{I}(\theta_1)}{(d-2)\delta^{d-2}}$. An important point to remember here is that we have used two different kinds of foliations for the HM volume ($u(t)$) and the volume between island and the left brane ($u(\mu)$). However, similar to the length of the HM and island surfaces, the divergence of the volumes for the two cases also cancel each other. Therefore, at least for our case, it seems that different foliations do not play a very nontrivial role in this picture. Also, when we look at the plots for the two volumes with time in Figure \ref{voltleftd4}, we find that the volume between the island and the left brane (constant) computed using this embedding is less than the HM volume at $t=0$. This is also something we expect ideally from Figure \ref{figislandvol} as pointed out.

	\subsubsection{General Discussion on Left Brane Volumes: }\label{covarguL}
	Here, we will argue how our computation of volume for left brane compares to the covariant maximal subregion complexity proposal. For the left brane, the HM surface is a time dependent surface. Therefore it is possible that the HM volume we computed in this paper (let's call this $V_{L-HM}(t)$) is not the maximal one ($V_{L-HM\left(\text{max}\right)}(t)$). However the maximal HM volume will always folow the relation $V_{L-HM\left(\text{max}\right)}(t)\geq V_{L-HM}(t)$. \par 
	
	However, the island surface is a time independent surface. In this case, the volume between the island surface and the left brane also is a time independent volume. Hence, Eq. \eqref{maxvol} reduces to Eq. \eqref{subcomp} for the island volume $\left(V_{L-IS}\left(t\right)=V_{L-IS}\left(t\right)=\text{constant}\right)$. Now we know that the island volume $V_{L-IS}$ for the left brane is less than HM volume $V_{L-HM}(t)$ at $t=0$. We also know that $V_{L-HM}(t)$ grows with time. Hence, the following inequality is followed for all times,
	
	\begin{equation}\label{covariantleft}
	V_{L-HM\left(\text{max}\right)}(t)\geq V_{L-HM}(t)> V_{L-IS}.
	\end{equation}

	Eq. \eqref{covariantleft} ensures the fact that at Page time the complexity jumps at a smaller value even if one follows the covariant subregion complexity proposal. Then for the left brane we are good with the complexity growing initially jumping to smaller constant value at Page time. 
	
	\subsection{Volumes Corresponding to Right Brane (Radiation):}
	In this section we follow the same protocol as described in the previous section with the modification that now we consider region between HM (/island) surface and the right brane. The right brane mimics the role of the non-gravitational bath in this model and therefore this volume is supposed to provide us with some understanding about how the complexity of the mixed radiation state evolves with time. From the previous study of left brane volumes, we expect the opposite phenomenon in case of the right brane volume, i.e., a jump at Page time. 
	\subsubsection{Hartman-Maldacena Volume:}

	\begin{figure}
		\centering
		\begin{subfigure}[b]{0.40\textwidth}
			\centering
			\includegraphics[width=\textwidth]{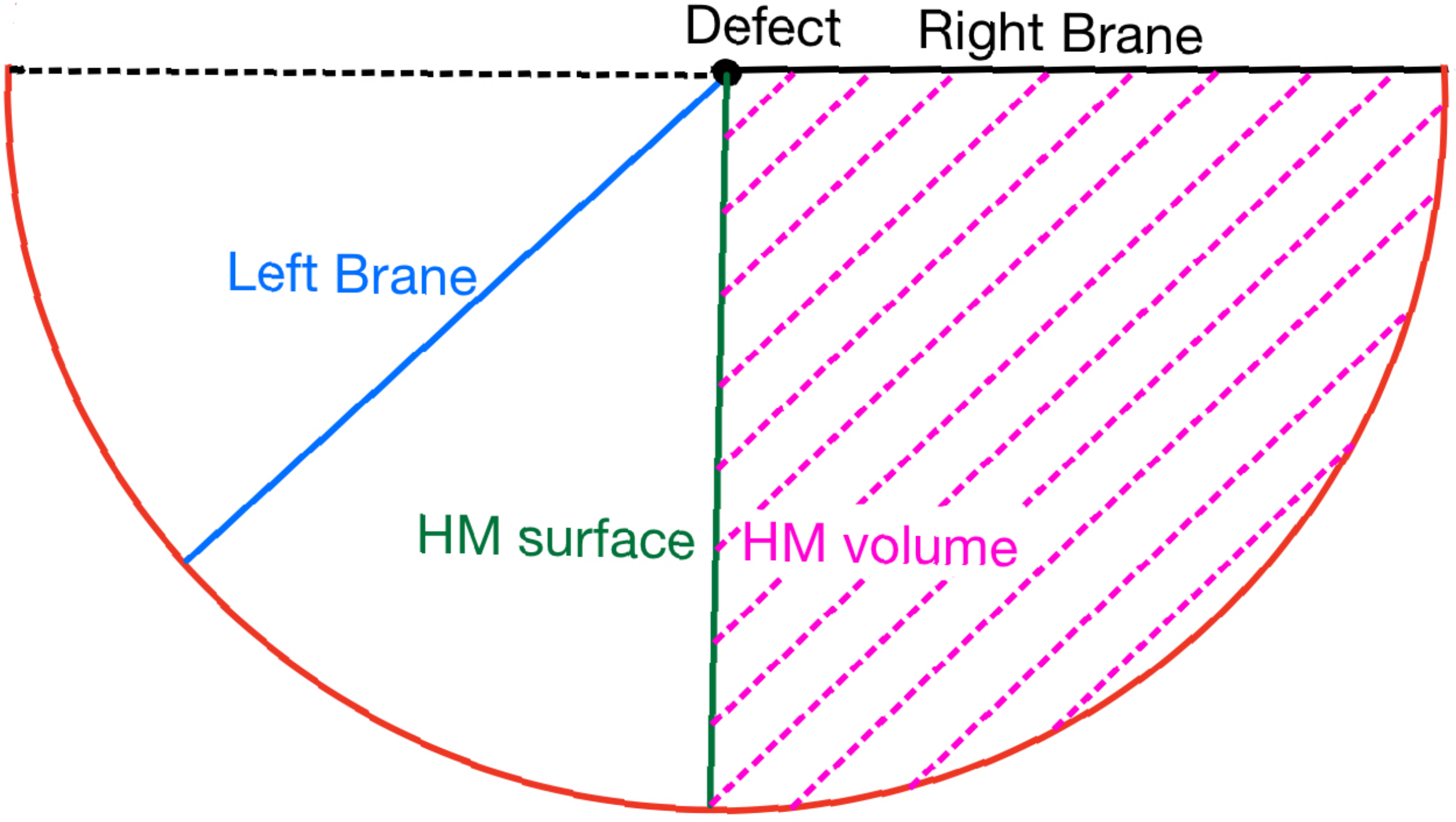}
			\caption{Volume between HM surface and right brane.}
			\label{volumepagerightHM}
		\end{subfigure}
		~~~~~~
		\begin{subfigure}[b]{0.39\textwidth}
			\centering
			\includegraphics[width=\textwidth]{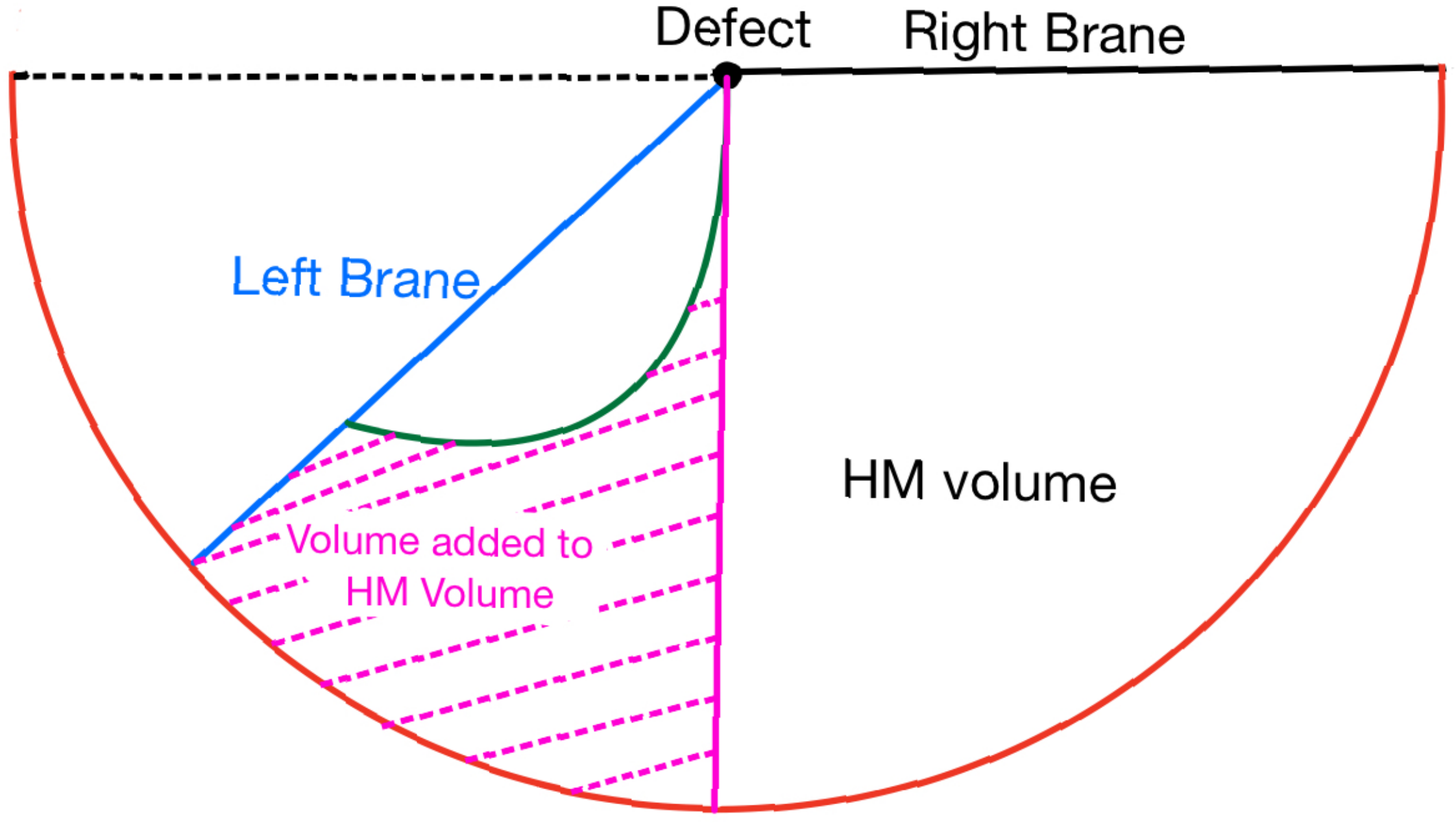}
			\caption{Volume added to right brane after Page time.}
			\label{figislandvolright}
		\end{subfigure}
		\caption{Volumes on the right brane corresponding to different RT surfaces are shown. Before Page time, it is the HM volume as shown in \ref{volumepagerightHM}. After Page time, a new bulk region gets added to the entanglement wedge of the right brane. This new region is contributes to the jump of volume at Page time and is shown in \ref{figislandvolright}.}
		\label{fig:rightvols}
	\end{figure}

	Until Page time, the right brane accesses the volume shown in Figure \ref{volumepagerightHM}. This region enclosed by the HM surface and the right brane also includes interior of black hole horizon and expected to grow with time similar to the area of the HM surface. We call this volume the HM volume for right brane. \par
	This part goes same as HM volume computation for the left brane. 
	We take the AdS$_{d+1}$ black string metric (mentioned in Eqs. \eqref{HMmetric} and \eqref{timeslicee}) and following the same argument as in subsection \ref{lefthmvoll}, we find the volume between the HM surface and right brane,
	\begin{align}
	V_{R-HM}(t_{\text{diff}})&=\mathcal{I}^{d} A^{\mathrm{reg}}_{HM}(t_{\text{diff}})\\
	\mathcal{I}^d&=\lim_{\delta \rightarrow 0}\int_{\frac{\pi}{2}}^{{\pi-\delta}}\frac{d\mu}{\sin^d\mu}\approx\frac{\delta^{1-d}}{d-1}+\mathcal{O}(\delta^{3-d})\\
	A^{\mathrm{reg}}_{HM}(t_{\text{diff}})&=\lim_{\delta\rightarrow{0}}\left[-\frac{1}{(d-2)\delta^{d-2}}+\int_{\delta}^{u_s}\frac{du}{|\dot{u}|u^{d-1}}\sqrt{-h(u)+\frac{\dot{u}^2}{h(u)}}\right]
	\end{align}
	The factor $\mathcal{I}^d$ is a purely divergent quantity by itself therefore at the end $V_{R-HM}$ will be divergent\footnote{This factor becomes finite if we take the right brane in the bulk and not in the conformal boundary. In that case, the model remains valid only if we fix the right brane, which is also gravitational, to be the gravitating bath region instead of considering dynamical bath region as done in \cite{Geng:2020fxl}.}. But our motivation is to capture the growth of $V_{R-HM}$ at late time hence the value of $\mathcal{I}^d$ is not essential.
	
	Now, we know that $A^{\mathrm{reg}}_{HM}$ is a monotonically increasing function of $t_{\text{diff}}$. Hence, asymptotically the volume $V_{R-HM}$ also goes linearly with $t_{\text{diff}}$.

	\begin{figure}
		\centering
		\includegraphics[scale=0.50]{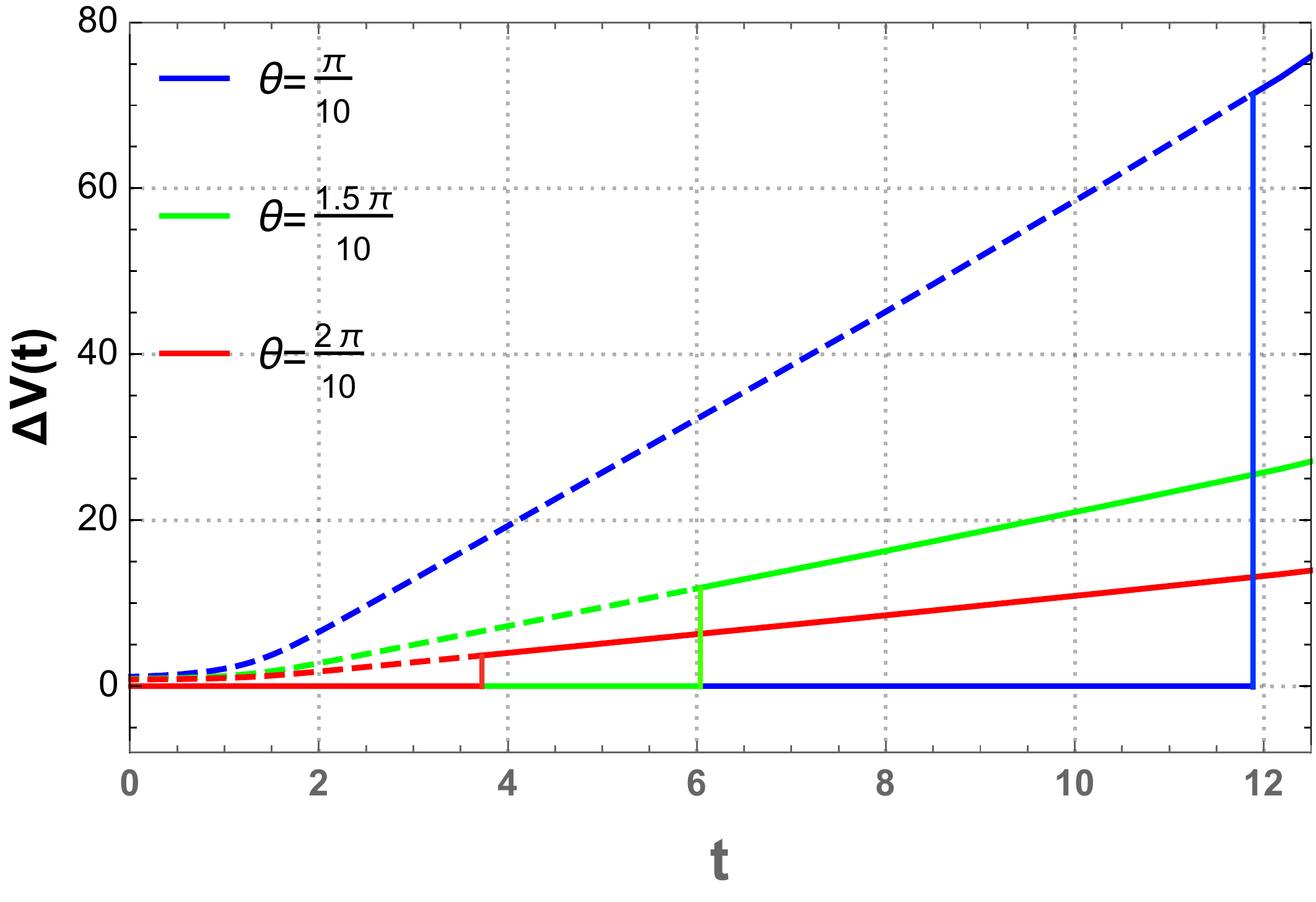}
		\caption{Volume vs time plot corresponding to right brane for $d=4$. As mentioned in the main text, we take the HM volume as the reference in this plot. Until the Page time, subregion complexity curve follows the HM volume and therefore it is zero in the plot. At Page time, for all the three physical brane angles, there is a volume that gets added due to the islands and this new volume grows with time.}
		\label{rightbranevol}
	\end{figure} 
	
	\subsubsection{Island Volume:}
	The volume between the island surface and right brane after Page time is denoted by the island volume. This region encloses the HM volume as well as a new volume bounded by the island surface, part of the left brane and the HM surface. This newly added region (shown in Figure \ref{figislandvolright}) also includes a part of the black hole interior and hence, we expect the change of volume at Page time to grow with time. This in general would mean that the slope of the complexity growth curve increases after the Page time. \par
	The computation of volume between the island surface and the right brane is tricky. We cannot foliate the the whole region between the island surface and the right brane by island surfaces. Thus we resolve this issue by first computing the volume between the Island and HM surface and then adding the volume $V_{R-HM}$ to it.
	\begin{equation}
	V_{R-IS}=(\mathcal{I}(\theta_1)+\mathcal{I}^d)\, \mathcal{A}_{HM}^{\mathrm{reg}}(t)-V_{L-IS}
	\end{equation}
	We define the difference between the two volumes as,
	\begin{equation}
	\Delta V(t)=V_{R-IS} \, \Theta(t-t_{\text{Page}})+V_{R-HM} \, \Theta(t_{\text{Page}}-t)-V_{R-HM}=(\mathcal{I}(\theta_1) \, \mathcal{A}^{\mathrm{reg}}_{HM}(t)-V_{L-IS}) \, \Theta(t-t_{\mathrm{Page}})
	\end{equation}
	The plot between $\Delta V(t)$ and $t$ is given shown in Figure \ref{rightbranevol}.

	It is important to note that $\Delta V$ is the difference between the chosen volume at a particular time and the volume between the HM surface and the right brane. Before Page time, this difference is always zero since HM surface is the preferred RT surface. Since we plot the difference $\Delta V$ remains zero until Page time. However, in fact, this reference volume between the HM surface and the right brane itself keeps growing linearly with time. Our reference volume is therefore itself a growing volume. After the Page time, island surface is the preferred RT surface and a volume gets added to the volume corresponding to right brane. This newly added volume makes $\Delta V$ nonzero starting from Page time. As we can see in Figure \ref{rightbranevol}, there is a jump in $y$ axis at the Page time. Another important point to note here is that the volume being added at Page time also grows linearly manifest through our plot. Therefore, had we plotted the actual volumes throughout instead of taking the primarily growing HM volume as the reference volume, the slope of the overall volume plot would increase at Page time. The jump will be unchanged however. 
	
	\begin{figure}
		\centering
		\includegraphics[scale=0.35]{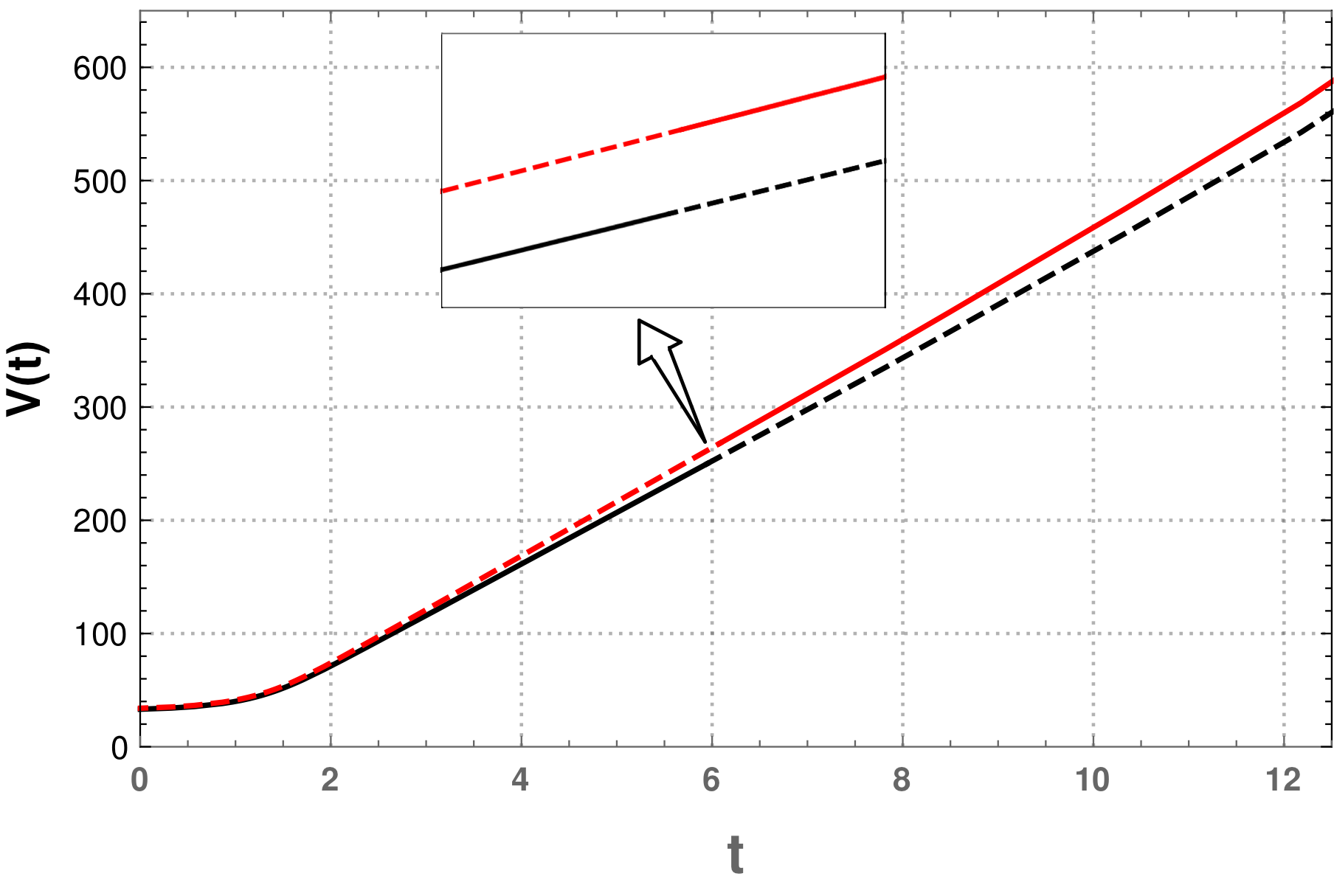}
		\caption{Volume plot for the case where fixed radiation is taken in the bulk. The black curve shows the HM volume whereas the red curve denotes the volume including the newly added volume (jump shown in the inset).}
		\label{rightbranefullvol}
	\end{figure}
	
	This complete volume can be studied if we take the right brane also in the bulk and fix the right brane to be the radiation region instead of considering dynamical bath. Let's say, we consider the case $d=4$ with left brane angle $\theta_{1}=1.5\frac{\pi}{10}$ and right brane angle $\frac{19 \pi}{20}$ (in the language of \cite{Geng:2020fxl}, the right brane angle is $\theta_{2}=\pi-\frac{19\pi}{20}=\frac{\pi}{20}<\theta_{1}$). In this case, the HM and island surface area remains the same. Hence, the entanglement plot remains the same as shown in blue curve of Figure \ref{areatd4}. Even, the volume plot corresponding to the left brane remains the same. But the HM volume corresponding to the right brane changes and there is no divergence as the right brane is also in the bulk now. Another way of saying this is simply that we have fixed the $\delta$ in Eq. $(29)$ to be $\frac{\pi}{20}$. In that case, the volume plot for the right brane becomes as shown in Figure \ref{rightbranefullvol}. In this plot, we can see that the slope (rate of increase) of the complexity increases after the jump at Page time. 
	
	\subsubsection{General Discussion on Right Brane Volume:}\label{covarguR}
	
	For the right brane the comparison between the boundary time volume and the covariant volume prescription gets a little complicated. The main reason of this complication is that even the island volume $V_{R-IS}(t)$ for right brane is time-dependent. Hence if we want to argue that the complexity jumping to higher value at Page time persists to the proposal in Eq. \eqref{maxvol}, we need to somehow argue that $V_{R-IS}(t)>V_{R-HM\left(\text{max}\right)}(t)$. What we only know so far is $V_{R-IS}(t)>V_{R-HM}(t)$ and $V_{R-HM\left(\text{max}\right)}(t)\geq V_{R-HM}(t)$ which is not enough to say 
	\begin{equation}\label{argright}
	V_{R-IS}(t)>V_{R-HM\left(\text{max}\right)}(t),
	\end{equation} 
	for $t\geq t_{\mathrm{Page}}$.\footnote{We are thankful to Hao Geng for pointing out this subtletly regarding the comparison with the covariant prescription.}
	
	In \cite{Hernandez:2020nem}, the authors followed the covariant proposal for subregion complexity.We expect the qualitative nature of the left brane volume to be proposal independent. For the right brane volume, the jump we found using our proposal is similar to the jump found in \cite{Hernandez:2020nem} from no island to the island phase which is also supports our findings. However, let us try to argue why in general we expect the radiation complexity to go through a jump at Page time logically. This has to do with the fact that the entanglement wedge (EW) of the radiation subsystem gets bigger suddenly at Page time as the entanglement islands become accessible to the radiation subregion. Hence, at any instant after Page time, for the right brane, the EW of the island phase contains the EW of the HM phase. Hence the maximal volume Cauchy slice of the former must also be bigger than the maximal volume Cauchy slice of the later. Therefore, the complexity in the HM phase is smaller than the complexity of the island phase starting from the Page time. This argument logically supports Eq.\eqref{argright}. Building on these arguments, we think that our plots for right brane is also universal and proposal independent and since the exact calculation is very hard to do in general, our example calculation is a good playground to understand the overall nature of the evolution of complexity.
	
	\section{Conclusion and Outlook:}\label{Conclusion}
	In this paper, we have worked with a KR brane model in general spacetime dimensions. The plots, however, has been shown explicitly for $d=4$. The bulk metric is taken to be that of the black string. The entanglement between the right and the left modes of the defect in this model follows a Page curve typical of the entanglement between eternal black hole and radiation, given the radiation is stored in a non-gravitational bath. From our study of volumes dual to the left and right modes, we look at the subregion complexity of the evolving black hole and radiation states, respectively. The entanglement islands play a crucial role in this study of volumes as well. However, we are far from fully understanding the effect of the islands in reproducing a Page curve of entanglement. In this particular model, we placed the gravitational brane, which we call the physical or left brane interchangeably, at a constant angle with the conformal boundary. The right brane is kept non-gravitational at the conformal boundary. However, one would still get the Page curve if one takes the right brane to be gravitational and computes the entanglement between the defect's left and right modes, given one fixes the right brane to be the region where the radiation is stored. As advocated in \cite{Geng:2020fxl}, there is a competition between the two brane angles with the conformal boundary. The entanglement curve will cease to exist if any of the two angles exceeds a particular dimension dependent angle, known as the critical angle ($\theta_{C}$). With this information in mind, we list the main findings of our paper in the following.
	
	\subsection*{Entanglement and the left brane volumes:}
	
	\begin{itemize}
		\item We apply Alishahiha's proposal of subregion volume complexity \cite{Alishahiha:2015rta} in our case and find the volumes on both the left and right brane corresponding to different preferred RT surfaces at different times. For the left brane, the volume provides the evolution of complexity for the eternal black hole. In contrast, for the right brane, the volume is argued to be dual to the evolution of the radiation state's complexity. 
		\item For the left brane, we find that the volume initially grows with time, similar to the Hartman-Maldacena (HM) surface, which is preferred until Page time. But unlike the HM surface, the HM volume depends upon the angle the gravitational brane makes with the conformal boundary. It is a very expected result since the volume enclosed by the HM surface and the left brane should, of course, depend upon where the brane is placed. We computed such volumes for three different constant brane angles $\left(\theta_{1}=\frac{\pi}{10}, \frac{1.5 \pi}{10}\, \, \text{and}\, \, \frac{2\pi}{10} \right)$ for $d=3, 4$ (these angles are less than the critical angle for both the dimensions) and found that the volume decreases as we increase the brane angle.
		\item Starting from the Page time, the island surface is the preferred RT surface. The island surface is specified by figuring out the critical anchor. We find the critical anchor by looking at the nature of the embedding. The basic idea here is that we have to pick such a boundary condition for which the surface can reach the defect ($\mu=\frac{\pi}{2}$) but can not reach the conformal boundary ($\mu=\pi$). The island surface has a time-independent surface area and therefore saturates the Page curve starting from the time when the area of the HM surface becomes greater than the island surface. These details of the critical anchor are not presented in any previous literature to the best of our knowledge.
		\item The volume corresponding to the island surface on the left brane is also a time-independent constant for a constant brane angle. On top of it, it is always less than the HM volume on the left brane at $t=0$. It is easy to understand this fact from the figures. The island volume remains constant because it is nowhere near the horizon of the black string metric. Therefore the volume between the island surface and the left brane does not go through any growth. We will see, however, that it is not the case for the right brane since in that case, the volume will also include behind-the-horizon regions.
		\item For the island volume, we find that the constant volume decreases as the physical brane angle is increased. The overall volume plot for the left brane for all the brane angles studied in this paper goes through a similar initially growing region, which goes through a negative jump at Page time and remains constant throughout the future. 
		\item In section \ref{covarguL}, we have provided the arguments to support the fact that this behaviour should persist even if one applies the ideal covariant prescription of the subregion complexity. This ensures that our results are qualitatively proposal independent. At least within this particular model's purview, the subregion complexity of the eternal BH should indeed follow the curve as shown in Figure \ref{voltleftd4}.

	\end{itemize}

	\subsection*{Right brane volumes:}
	\begin{itemize}
		\item For the right brane, the situation is a bit different. The HM volume on the right brane is also a growing one. However, we find that since we take the right brane at the conformal boundary, the HM volume has a nontrivial divergence, regulating which would involve a cut-off. We do not choose this constant arbitrarily. Rather, we treat this growing volume as our reference volume for all the times and compute the change of volume ($\Delta V$) from this growing volume as we move forward in time.
		\item After Page time, the island surface comes into the picture. This gives the right brane access to a new volume bounded by the HM surface, the island surface and the left brane. Hence, $\Delta V$ goes through a positive jump at Page time. Afterwards, this change keeps growing with time, indicating that the newly accessible volume also includes the behind-the-horizon region. This is unlike the volume corresponding to the island surface for the left brane, which is constant. The fact that this newly added volume decreases as we increase the physical brane angle remains unchanged for the right brane.
		\item Although we did only compute the change of volume from the reference volume and the reference volume growth is not shown for the right brane at the conformal boundary. It can be studied if we take the right brane to be gravitational and keep the brane angle less than the physical brane angle. In that case, one would explicitly find that the slope of the complexity growth increases at Page time with the positive jump as shown in Figure \ref{rightbranefullvol}. Therefore, the general fact for the right brane from our finding is that the complexity keeps growing with time, even after Page time. The slope of the growth increases due to the inclusion of the islands, and the jump happens at the inclusion point (Page time).
		\item The entanglement curve and the left brane volume plot remain unchanged even when we take the right brane in the bulk until we make sure that $\theta_{1}>\theta_{2}$. 
		\item Finally, similar to the left branes, we have again argued for our results being valid qualitatively even for the maximal subregion complexity proposal by providing a set of logical explanations in section \ref{covarguR}. 
	\end{itemize}
	
	Building on the main findings, we now try to explain them in terms of our present understanding and results presented in other literature on the related issue. The jump in volume at Page time for the right brane is not new and has been previously found in \cite{Ben-Ami:2016qex, Bhattacharya:2020uun, Hernandez:2020nem}. It is typical for the radiation state as the islands are included in the entanglement wedge of the radiation subsystem starting from the Page time. This results in the first-order phase transition of the volume curve at Page time. This has been related to the multipartite purification complexity of the Hawking quanta in \cite{Bhattacharya:2020uun, Bhattacharya:2020ymw}, whereas the authors of \cite{Hernandez:2020nem} have argued it to be related to the mutual complexity. 
	
	The complexity of purification is the complexity that gets added to the previous complexity of the radiation state due to the purification of certain hawking quanta outside the black hole horizon with their inner partner modes accessible to the radiation subsystem at the Page time. This newly accessible modes purify the previously present partner modes and to construct these purified partner modes, one needs to use certain new number of quantum gates from the perspective of the radiation side. This causes the jump in volume at Page time. From the black hole side, the situation is just the other way around. Due to the monogamy property of entanglement, the black hole can not access the purified modes anymore and needs a lesser number of gates than before Page time to reproduce the state in the black hole side. However, it is hard to understand completely why the left brane volumes become time-independent starting from the Page time. Physically it indicates the black hole needs very small number of gates compared to before Page time to construct the mixed state accessible to itself. Since this time independence results from the island surface's time independence (remember this time independence helps the entanglement curve achieve the saturation), it seems that the constant entanglement results in the constant complexity of the eternal black hole state. It will nevertheless be interesting to study other models of the eternal black hole-radiation system to determine if this behaviour is universal or model-dependent. However, we would like to stress that the positive (negative) jump for the radiation (eternal black hole) complexity at Page time seems universal and model-independent. More arguments on this particular direction can be found in the conclusion section of \cite{Bhattacharya:2020uun}.
	
	In \cite{Hernandez:2020nem}, the authors have attributed the jump in complexity (on the brane) from no-island to island phase to the mutual complexity. For two sub-region $A$ and $B$, mutual complexity is defined as $C_{\mathrm{mutual}}(AB)=C(A)+C(B)-C(A\cup B)$. Since in \cite{Hernandez:2020nem}, the RT surfaces were disconnected after Page time, the argument of mutual complexity seemed to work. However, in our case, the mutual complexity ideally should be zero since the island surface is connected and divides the region into two bulk volumes the sum of which give us the complete volume. Therefore, the connection to purification complexity seems to work better in our case. In general, the subregion complexity is expected to be related to purification complexity, which can be related to mutual complexity (as advocated in \cite{Camargo:2018eof}). It would be interesting to look more into the connection of purification and mutual complexity in this regard. One might hope to find such a relation indeed if the covariant maximal volume prescription is followed. 
	
	Finally, to conclude, we would like to mention that ideas of complexity in the field theory side are still developed partially \cite{Jefferson,Chapman:2017rqy,Hackl:2018ptj,Khan:2018rzm,Bhattacharyya:2018bbv,Ali:2018fcz, Caputa:2017urj, Caputa:2017yrh, Bhattacharyya:2018wym}\footnote{Again, the list is by no means complete and curious readers are encouraged to look into the references and citations of these papers.}. Hence, we followed the gravitational (holographic) mixed state complexity proposal previously explored in various literature \cite{Alishahiha:2015rta, Bhattacharya:2019zkb, Abt:2017pmf, Abt:2018ywl, Geng:2019ruz, Geng:2019yxo, Banerjee:2017qti, Banerjee:2020gyv, Bhattacharya:2020uun} and applied in the model studied in this paper. The results indicate a universality in the jump of radiation complexity at Page time and provide us with some hope that the path being followed is somewhat self-consistent. However, to complete the understanding, it would be extremely worthwhile to apply the field-theoretic proposals of complexity within simpler settings. One possible way is to apply the ideas of purification complexity developed in the field theory and CFT sides to study a partial purification of an interacting system at the midpoint. Since in the field-theoretic side, particularly for CFTs, people are still exploring very specific states, it might be hard to address complicated mixed states in CFTs. But, applying it to free field theoretic models might also teach us some important insights about this particular phenomenon, and its dependence on the correlation strength of the field theory. Since the field-theoretic studies of mixed state complexity have also made forays in probing other interesting physics like chaos, quench, etc \cite{Ali:2018aon, Camargo:2018eof, Ali:2019zcj, Bhattacharyya:2019txx, Yang:2019iav,Bhattacharyya:2020art,Bhattacharyya:2020iic,Balasubramanian:2021mxo}, it would also be immensely exciting if one could tie the ideas of islands and information paradox with scrambling or other interesting physics happening inside the black holes. Holographically, it also remains an open problem to build a consistent model-independent quantitative understanding of this jump at Page time. The dependence on the theory's parameters and, more precisely, having better explicit control on the rate of radiation, Page time etc., seem to be the crucial direction one probably could also be interested in. On the other hand, to check whether the overall nature of the plots is universal, it might be a good idea to study complexity in other models of evaporating (for example; holographic moving mirror \cite{Akal:2020twv, Reyes:2021npy}) or eternal (for example; geometric secret sharing \cite{Balasubramanian:2020hfs}) black hole vs radiation. We hope to explore these directions more in future.  
	
	\section*{Acknowledgements}
	
	The authors would like to thank Hao Geng for collaboration in the early stages of this work and comments on the draft. A.B.$(1)$ would like to thank Shibaji Roy, Christian Northe, Sabyasachi Maulik and Arnab Kundu for discussions on related issues. P.N. thanks Budhaditya Bhattacharjee and Pingal Pratyush Nath for useful discussions. A.B.$(2)$ is supported by  Start Up Research Grant (SRG/2020/001380) by Department of Science \& Technology Science and Engineering Research Board (India). P.N. acknowledges University Grants Commission (UGC), Government of India for providing financial support. A.K.P. is supported by the Council of Scientific \& Industrial Research (CSIR) Fellowship No. $09/489(0108)/2017$-EMR-I.
	
	\appendix
	\section{Appendices}
	\subsection{Area and Volume Regularization:}\label{appa}

	Both the area of the HM and Island surfaces diverge as they reach the defect at $\mu=\frac{\pi}{2}$. There are many ways to regularize these divergences. The most usual way followed in many of the literatures is to introduce a UV cut off. However, this is best applied in cases where the computations are analytic and the contribution can be cancelled by subtracting the empty AdS divergence. In our case, the solutions of the embedding for both the HM and the island surfaces are numerical and hence, considering a simple UV cut off might be problematic at times. Therefore, we initially resolved to a different path. We know that the HM surface always orthogonally connects the two defects and is placed precisely at $\mu=\frac{\pi}{2}$. It grows with time. But the divergence does not grow as the divergence is there simply due to the physics near the defect which has nothing to do with the horizon of the black string metric. The growth of HM surface on the other hand is a result of the black string metric. So in case of the HM surface, we can simply subtract the divergence of the empty AdS HM divergence from the black string HM divergence to get the finite contribution in area.

	\begin{figure}
		\centering
		\includegraphics[scale=0.4]{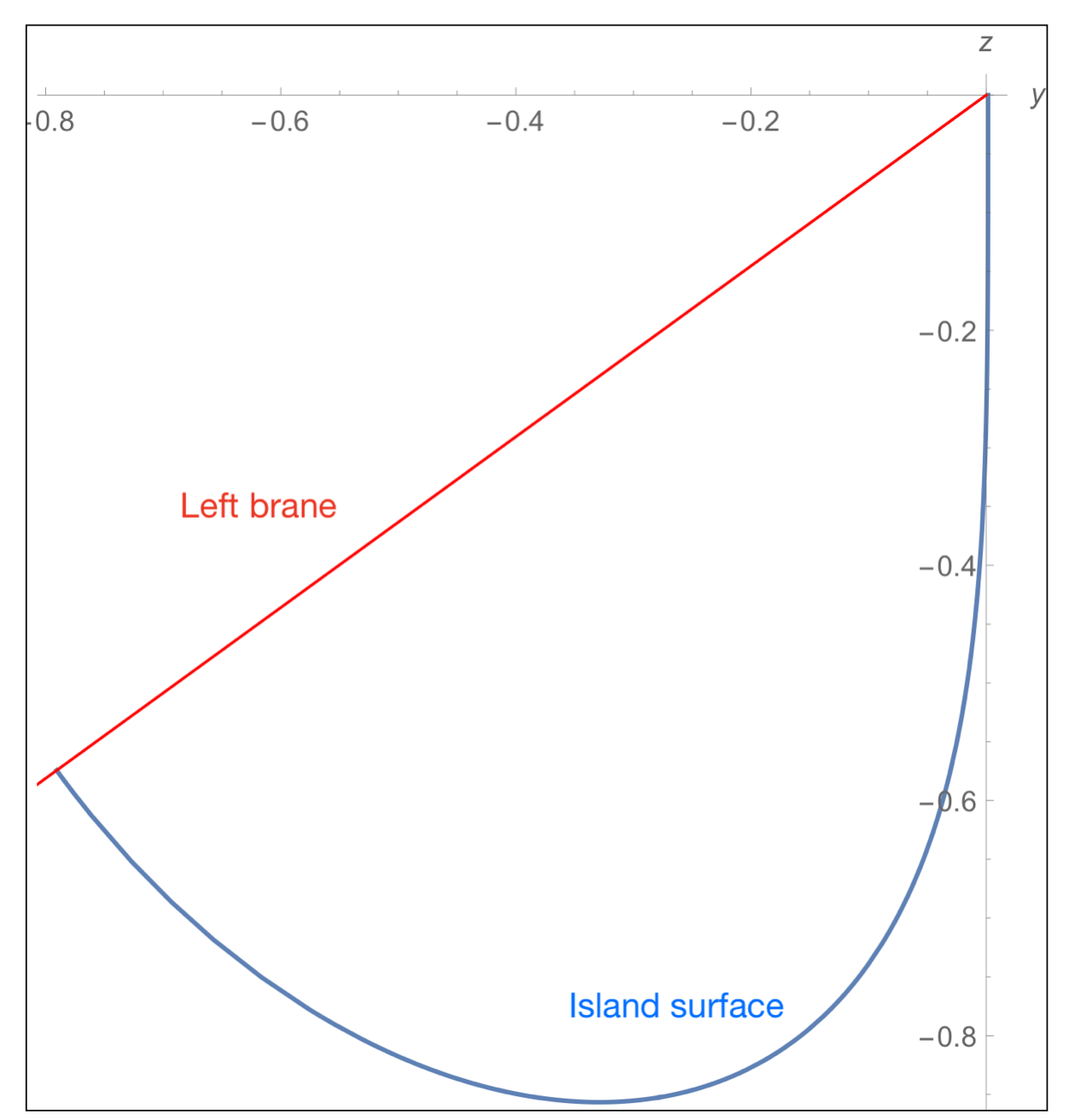} 
		\caption{The blue line is the island surface that meets the left brane marked by the red line. The island surface can be seen to meet the vertical line (the HM surface) at around $u=0.3$. The coordinates are defined as $z=u \sin \mu$ and $y=-u \cos \mu$ \cite{Geng:2020fxl}.}
		\label{HMmeetisland}
	\end{figure}

	In case of the island surface, the situation is bit trickier. As we mentioned previously, the island surface is specified by looking at the critical anchor plots of the time independent embedding function $u[\mu]$. But, for the critical anchor choice defining the island, we find that the area diverges numerically when we input the embedding function with the proper boundary conditions. In this case, it is even harder to subtract a UV divergence. We plot the embedding function with the $\mu=\frac{\pi}{2}$ line (remember that the HM surface is placed along this line). We find that the island surface starts getting overlapped completely with the HM line from certain $u$ value. We use this fact subsequently to subtract the HM divergence from the divergent island contribution starting from the region where these two surfaces start becoming the same. This assumption of considering the two surfaces to be the same starting from some $u$ value is therefore crucial in cancelling the divergence, especially of the island surfaces. A representative figure of this fact is shown in Figure \ref{HMmeetisland}.
	
	To make the story analytically consistent as well, the divergence in Island area can also be shown to exactly cancel the divergence coming from the HM surface area by using the fact that $\mu\rightarrow\frac{\pi}{2}$ and $\mu'(u)\rightarrow 0$ as $u\rightarrow 0$. To see that we express the Island area as an $u$ integral,
	\begin{eqnarray}
	\mathcal{A}_{IS}&=&\int\frac{d\mu}{u^{d-2}(\sin\mu)^{d-1}}\sqrt{1+\frac{u'(\mu)^2}{u^2h(u)}}\\
	&=&\int\frac{du}{u^{d-2}\sin^{d-1}\mu}\sqrt{\mu'(u)^2+\frac{1}{u^2 h(u)}}
	\end{eqnarray}
	Hence the divergence in the island area when $u= \delta\rightarrow 0$ is $\frac{1}{(d-2)\delta^{d-2}}$, which is same as the cutoff dependence of the HM area.
	
	We apply similar idea while regularizing the volume as well. This is briefly explained previously in Section \ref{volleft}.
	
	\begin{figure}
		\centering
		\includegraphics[scale=0.5]{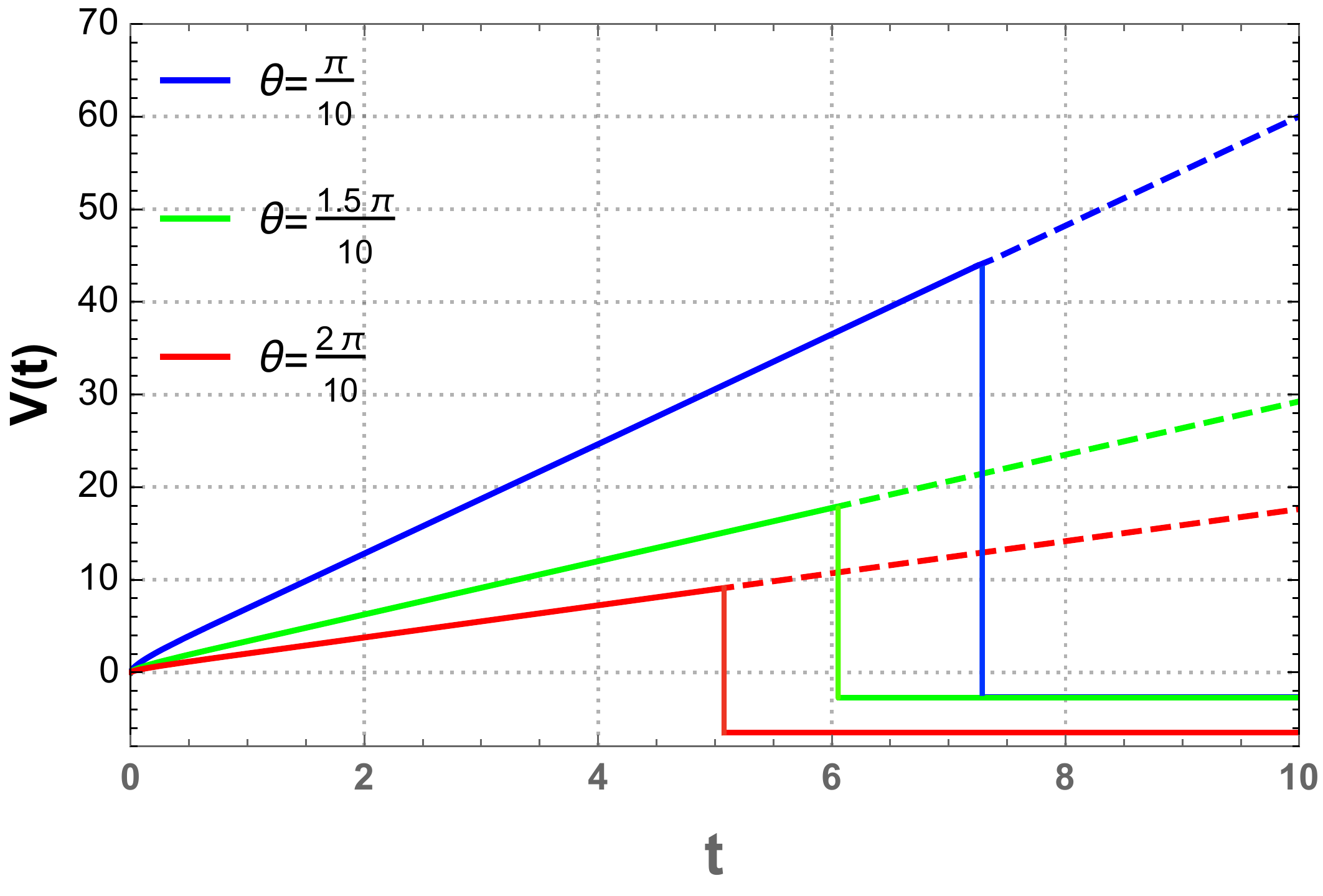} 
		\caption{Volume vs time plot corresponding to left brane for $d=3$. The finite part of the HM volume is positive whereas the finite part island volume is negative. }
		\label{volvstimed333}
	\end{figure}
	
	\subsection{Volume Negativity:}\label{appb}
	
	We find that the finite part of the volume (after subtracting the otherwise divergent part) between the island and the left brane could be negative. In $d=3$ we find that this volume is always negative irrespective of the brane angle but in $d=4$ this volume is only negative if $u_1$ is less than some critical value. For $d=3$, it is somewhat expected. The finite part of HM volume is just a scale factor multiplied by the HM area, which starts from zero for $d=3$. We also know that the island volume which is constant throughout the time is expected to be less than HM at $t=0$ and hence less than zero. We would like remind here that the crucial point is that this does not mean the volume we compute is unphysically negative. It is only the finite part after subtracting the divergent part. Had we not done so, the overall volume would have of course been a very large and positive value. The volume plot corresponding to the left brane is shown in Figure \ref{volvstimed333}.

\bibliographystyle{JHEP}
\providecommand{\href}[2]{#2}\begingroup\raggedright\endgroup

\end{document}